\documentclass[aps,prr,twocolumn,superscriptaddress,10pt]{revtex4-1}
\usepackage[colorlinks ,linkcolor=blue,anchorcolor=blue,citecolor=blue,urlcolor=blue]{hyperref}
\usepackage{amsmath}
\usepackage{amssymb} 
\usepackage{booktabs}
\usepackage{multirow}
\usepackage{mathtools}
\usepackage{physics} 
\usepackage{bm}
\usepackage{graphicx}
\usepackage{epstopdf}
\usepackage{tikz}
\usepackage{subfigure}
\usepackage{float} 
\usepackage{placeins} 
\usepackage{flushend,lipsum} 
\usepackage{color}

\usepackage[normalem]{ulem}
\usepackage{algorithmicx,algpseudocode}

\def\tr{\mbox{tr}}

\begin{document}
\title{Enhancing Circuit Fidelity in Transmon Qubit Rings via Operation Duration Tuning under Strong Connectivity Noise}
\author{Quan Fu}
\affiliation{School of Physics and Technology, Wuhan University, Wuhan
	430072, China}
\affiliation{Quantum Science Center of Guangdong-Hong Kong-Macao Greater Bay Area, Shenzhen, Guangdong 518045, China}
\affiliation{Department of Physics, City University of Hong Kong, Tat Chee Avenue, Kowloon, Hong Kong SAR, China}
\author{Xin Wang}
\affiliation{Department of Physics, City University of Hong Kong, Tat Chee Avenue, Kowloon, Hong Kong SAR, China}
\affiliation{City University of Hong Kong Shenzhen Research Institute, Shenzhen, Guangdong 518057, China}
\author{Rui Xiong}
\email{xiongrui@whu.edu.cn}
\affiliation{School of Physics and Technology, Wuhan University, Wuhan 430072, China}	
	
	\begin{abstract}
		Superconducting transmon qubits are a promising platform for quantum computation, yet they face significant fidelity degradation due to connectivity noise, particularly in the intermediate coupling regime where noise levels are substantial. While prior works suggest that high fidelity requires operating in regimes with strongly suppressed noise, maintaining such conditions under practical experimental constraints remains challenging. To address this, we investigate quantum gate operations in fully connected transmon rings, examining both SWAP and general circuits. Our study reveals that fidelity can be significantly enhanced by tuning gate operation durations, with local maxima emerging even under strong noise conditions. These fidelity enhancements occur consistently across different qubit numbers and operation types, and for specific initial states—particularly those with favorable symmetry or entanglement properties—the achieved fidelities approach quantum error correction thresholds. Furthermore, we develop a supervised machine learning model that accurately predicts the optimal operation durations for new devices, enabling efficient optimization without extensive experimental simulations. These results provide a pathway toward robust quantum circuit design in noisy experimental environments.
		
	\end{abstract}
	
	\maketitle
	
	\maketitle
	
	\section{Introduction}
	
	Superconducting qubits represent a leading platform for scalable quantum computation, combining precise controllability with the potential for extended coherence under optimized conditions~\cite{Blais2024,Blais2007,Schuster2007}. Building on this foundation, these systems exploit Josephson junctions to enable rapid electrical manipulation of quantum states, forming the basis for gate-based architectures~\cite{Josephson1962,Josephson1974,You2011}. Among various implementations, the transmon qubit—a refined Cooper pair box design—has emerged as the dominant architecture due to its reduced charge noise sensitivity and enhanced coherence properties~\cite{Koch2007,Paik2011}. Typically fabricated from aluminum and operated below 30 mK, transmons achieve strong coupling to coplanar waveguide (CPW) resonators by concentrating electric fields near the qubit~\cite{Yoshihara2017,Kollar2019}. While this configuration enhances dispersive interactions vital for quantum information processing~\cite{Jaynes1963,Cummings2013}, it simultaneously introduces connectivity induced noise that significantly degrades fidelity in multiqubit operations~\cite{Robert2022,Fu2024}.
	
	This inherent trade-off between connectivity and noise leads to a critical challenge: although shorter gate durations minimize noise accumulation, they fundamentally limit the complexity of implementable quantum operations~\cite{Robert2020,Foulk2022}. To quantify this trade-off, consider the SWAP gate—a key primitive for state transfer—whose duration scales as $\tau = \pi/(4J)$ (with $\hbar=1$), where $J$ represents the qubit-qubit coupling strength. For a fixed noise amplitude $\lambda_0$, conventional wisdom suggests that infidelity should decrease monotonically with increasing $J/\lambda_0$, with high-fidelity operation requiring $J/\lambda_0 \gtrsim 100$ in transmon devices~\cite{Barends2014,Adam2022,Noiri2022}. However, maintaining such high ratios proves experimentally challenging in practical strong-coupling regimes~\cite{Caldwell2018,Kong2015}, particularly in the noisy intermediate-scale quantum era where precise control over environmental parameters is limited.
	
	To address this fundamental limitation, we demonstrate that optimal operation points emerge in the more accessible intermediate-coupling regime ($10 \lesssim J/\lambda_0 \lesssim 100$) for fully connected transmon rings~\cite{Deng2015,Kong2015,Zhu2017}. We define an optimal operation point as a specific gate duration (or equivalent $J/\lambda_0$ ratio) where gate fidelity reaches a local maximum despite substantial noise, representing a dynamic balance between operation time and decoherence. Crucially, this approach contrasts with the conventional strategy of simply minimizing gate duration and enables longer duration operations under realistic noise conditions. Our investigation begins with SWAP gate sequences and extends to general randomly generated operations~\cite{Robert2020,Foulk2022}, revealing that optimal points appear consistently across different systems with positions largely independent of the specific circuit implementation. Remarkably, for certain initial states—particularly GHZ-type entangled states—fidelity at these optimal points approaches the error-correction threshold of 99.9\%~\cite{Kandel2019,Sigillito2019}, highlighting the important role of state-circuit symmetry in noise resilience.
	
	While these optimal operation points offer significant advantages, their experimental implementation faces practical challenges due to device-to-device variations in CPW coupling strengths and noise distributions. To overcome this issue, we integrate a supervised machine learning approach that accurately predicts optimal operation points using limited simulation data~\cite{Czarnik2021,Strikis2021,Cincio2021}. This methodology avoids the need for exhaustive simulations for each parameter configuration~\cite{Zlokapa2020,Rodrigues2023} and facilitates adaptive optimization of gate durations across diverse experimental setups.
	
	The remainder of this paper is organized as follows: Section~\ref{sec:Model and method} details the model and method. Sections~\ref{sec:SWAP} and~\ref{sec:generalU} present comprehensive results for SWAP gates and general operations, respectively. Section~\ref{sec:states} explores the dependence of fidelity on initial states, Section~\ref{sec:mechanism} discusses the underlying physical mechanisms, and Section~\ref{sec:ML} describes the machine learning framework for optimal point prediction. Finally, Section~\ref{sec:conclusion} summarizes our key findings and discusses their implications for future quantum processor design.
	
	\section{Model and method}
	\label{sec:Model and method}
	
	\begin{figure}[t]
		\centering
		\includegraphics[width=1\linewidth]{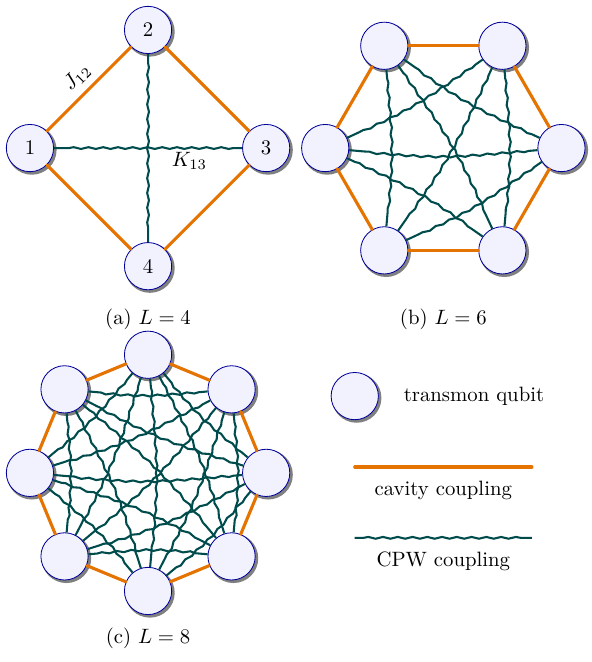}
		
		\caption{Schematic of transmon qubit devices with different numbers of qubits $L$.
				Each node represents a transmon qubit, with solid lines denoting nearest-neighbor (cavity-mediated) couplings and wavy lines denoting long-range (CPW-bus-mediated) couplings.
				Panel (a) highlights examples of a nearest-neighbor coupling $J_{12}$ and a remote CPW-bus coupling $K_{13}$.}
		\label{fig:connectivity}
	\end{figure}
	
	Transmon qubits are coupled through two primary mechanisms: nearest-neighbor couplings mediated by on-chip superconducting resonators and long-range couplings mediated by CPW transmission lines acting as quantum buses. In experimental devices, nearest-neighbor couplings are often implemented using dedicated tunable couplers (which are themselves transmon qubits) for compactness and fast control. In our theoretical model, we abstract this implementation detail and represent the nearest-neighbor exchange interaction as a cavity-mediated coupling with strength $J_{ij}(t)$. The long-range CPW-mediated couplings, represented by $K_{ij}(t)$, enable strong indirect interaction between spatially separated qubits and have been used in quantum bus architectures to facilitate long-range entanglement and quantum information transfer~\cite{Kollar2019}. While CPW transmission lines are physically larger than qubits, they provide a viable route to implement all-to-all connectivity in modular quantum processors. Our study focuses on the theoretical implications of such hybrid connectivity and its associated noise on circuit fidelity.
	
	We study three transmon qubit devices with varying qubit numbers $L$, as shown in Fig.~\ref{fig:connectivity}. Each vertex represents a transmon qubit, with solid lines indicating nearest-neighbor couplings and wavy lines indicating long-range CPW-bus-mediated couplings. The devices form ring structures where adjacent qubits are connected via nearest-neighbor couplings, while CPW links provide essential long-range interactions for quantum information distribution.
	
	The control Hamiltonian governing these systems is:
	\begin{equation}
		\begin{aligned}
			H_{\mathrm{ctrl}} &= \sum_i \left[ \Omega_i^x(t) \left( \cos\phi_i(t) \sigma_i^x + \sin\phi_i(t) \sigma_i^y \right) + \frac{\Delta_i(t)}{2} \sigma_i^z \right] \\
			&+ \sum_{\langle i,j\rangle} 2J_{ij}(t) \left( \sigma_i^+ \sigma_j^- + \mathrm{h.c.} \right)\\
			&+ \sum_{\{i,j\}} 2K_{ij}(t) \left( \sigma_i^+ \sigma_j^- + \mathrm{h.c.} \right),
		\end{aligned}
		\label{eq:H_ctrl}
	\end{equation}
	where $ \sigma_{i}^{x}, \sigma_{i}^{y}, \sigma_{i}^{z}$ are Pauli matrices for qubit $ i $, $\sigma_{i}^{\pm}$ are raising/lowering operators, $\Omega_i^x(t)$ is the single qubit drive amplitude, $\Delta_i(t)$ is the detuning, $\langle i,j\rangle$ denotes adjacent qubits coupled via a resonator or tunable coupler with strength $J_{ij}(t)$, and $\{i,j\}$ denotes non-adjacent qubits coupled via a CPW bus with strength $K_{ij}(t)$ [cf. Fig.~\ref{fig:connectivity}(a)]. Appendix~\ref{sec:full_model} provides the full derivation from circuit Hamiltonians.
	
	We note that in our model, the coupling strengths $J_{ij}(t)$ and $K_{ij}(t)$ are treated as tunable control parameters. In actual devices, these couplings are often adjusted by changing the qubit frequencies (thus modifying the detuning from a shared resonator or bus), which introduces correlations between different couplings. However, our theoretical abstraction allows us to focus on the generic effects of connectivity noise. For SWAP gate sequences, we consider sequential operations where only one coupling is active at a time, which can be achieved by detuning non-target qubits, thereby decoupling the couplings in time. This is consistent with the commonly used "activate/deactivate" control strategy in experiments. For general multi-qubit unitaries, the universal quantum computing theory ensures that any operation can be decomposed into single- and two-qubit gates, each of which can be realized by adjusting the couplings without fundamental limitations. Although correlations between couplings exist in actual devices, optimization algorithms that take these correlations into account can find pulse sequences to achieve the target operation. Our model, by assuming tunable control, captures the essential physics of connectivity noise and the existence of optimal operation points.
	
	This Hamiltonian enables universal quantum computation by generating any unitary $R \in \text{SU}(2^L)$~\cite{Lloyd1995,Dawson2006}. For SWAP gates, we set $\Omega_i^x(t)=\Delta_i(t)=K_{ij}\equiv 0$ and $J_{ij}=J$ for the target qubit pairs. For general operations $R$, we compute the time-dependent parameters using the procedure in Appendix~\ref{sec:Implementation}.
	
	\begin{figure}[t]
		\centering
		\includegraphics[width=0.5\linewidth]{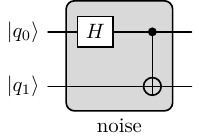}
		\caption{Schematic of a two qubit quantum circuit subject to connectivity noise.
			The circuit (Hadamard followed by CNOT) operates under noisy conditions, with shaded regions indicating noise affected segments.
			In transmon systems, this noise primarily stems from connectivity induced fluctuations.}
		\label{fig:qc_2qubit}
	\end{figure}
	
	\subsection{Connectivity noise model}
	
	The central challenge in achieving high fidelity operations arises from connectivity induced noise, which becomes particularly significant in the intermediate coupling regime ($10 \lesssim J/\lambda_0 \lesssim 100$). We focus on two dominant noise sources: parasitic capacitance noise in cavity coupled pairs and microwave photon loss in CPW channels~\cite{Kollar2019,Kong2015}. For a target unitary $R$, noisy processes transform the evolution to $U^{\dagger}R$, altering the final state as illustrated in Fig.~\ref{fig:qc_2qubit}.
	
	The noise Hamiltonian is:
	\begin{equation}
		\begin{aligned}
			H_{\mathrm{noise}}   &= \sum_{i}\delta\omega_{i}\sigma_{i}^{z} +\sum_{\langle i,j\rangle}  \lambda_{ij}^{J}(t)  \sigma_{i}^{y}\sigma_{j}^{y} \\
			&+  \sum_{\{i,j\}}  \lambda_{ij}^{K}(t) (\sigma_{i}^{+}\sigma_{j}^{-} + \sigma_{i}^{-}\sigma_{j}^{+}).
		\end{aligned}
	\end{equation}
	The cavity noise $\lambda_{ij}^{J}(t)$ originates from parasitic capacitance ($\lambda_{ij}^{J} \propto C_g^{\text{para}}$), while CPW noise $\lambda_{ij}^{K}(t)$ comes from photon loss ($\lambda^{K}_{ij}\propto Q^{-1/2}$)~\cite{Zhao2025,Zhang2019}. Fabrication variations yield characteristic ratios $\mathcal{R} = \lambda_{ij}^{K} / \lambda_{ij}^{J}$ between 3 and 20~\cite{Kong2018thesis}, which we maintain in simulations. This range is motivated by typical experimental values: parasitic capacitance noise $\lambda_{ij}^{J} \sim 0.05$--$0.5$ MHz and CPW photon-loss noise $\lambda_{ij}^{K} \sim 1$--$5$ MHz. Frequency shifts $\delta\omega_i$ are minor perturbations ($\delta\omega_i/\lambda_{ij}^{K} \sim 0.1\%$). The chosen values of $\delta\omega_i $ correspond to dephasing times $T_2^* \gtrsim 20\,\mu$s, achievable in state-of-the-art devices. This allows us to isolate connectivity-noise effects. Furthermore, our results exhibit scale-invariance: scaling all noise amplitudes by a factor $\alpha$ simply rescales the optimal gate duration as $\tau_{\text{opt}} \propto 1/\alpha$, leaving the fidelity profile versus $J/\lambda_0$ unchanged.
	
	We note that in a more detailed physical model, tuning the qubit frequencies (via $\Delta_i(t)$ or via design parameters $\Delta_{ik}$) could modify the coupling strengths $g_k^i(\omega)$ and consequently the noise amplitudes. However, in our quasi-static noise model, we treat $\lambda_{ij}^{K}$ and $\lambda_{ij}^{J}$ as fixed parameters determined by static device properties. This simplification is valid when the dynamic frequency shifts used for gate operations are small compared to typical qubit-resonator detunings. Our study focuses on the effect of fixed noise amplitudes on circuit fidelity as a function of gate duration; a more detailed noise model that includes dependence on operational parameters is an important direction for future device-specific optimization.
	
	Given the slow noise dynamics in transmon systems compared to gate operations~\cite{Koch2007,Robert2022}, we model quasistatic noise using normal distributions: $\lambda_{ij}^{K}(t) \sim \mathcal{N}(\lambda_{ij,0}^{K}, \sigma_{K}^2)$ and $\lambda_{ij}^{J}(t) \sim \mathcal{N}(\lambda_{ij,0}^{J}, \sigma_{J}^2)$~\cite{Fu2024}.
	
	\subsection{Fidelity metric and simulation parameters}
	
	We quantify operation quality using the state fidelity:
	\begin{equation}\label{eq:fidelity}
		F = \left| \langle\Psi_0 | R^\dagger U^{\dagger}R | \Psi_0 \rangle \right|^2,
	\end{equation}
	where $|\Psi_0\rangle$ is the initial state. For each parameter set, we generate 1,500 noise samples from Gaussian distributions and compute the mean fidelity $\overline{F}$. 
	
	Using $\lambda_0$ as the energy unit, we set $\lambda_{ij,0}^{K}=\lambda_0$, $\lambda_{ij,0}^{J}=0.1\lambda_0$, $\delta\omega_{i}=10^{-4}\lambda_0$, with relative fluctuations $\sigma_{K}=0.01\lambda_{ij,0}^{K}$ and $\sigma_{J}=0.01\lambda_{ij,0}^{J}$. These values align with experimental transmon devices (Appendix~\ref{sec:full_model}). The key parameter $J/\lambda_0$—which directly relates to gate duration through $\tau \propto 1/J$—is varied across the intermediate coupling regime to identify optimal operation points.
	
	We consider three classes of initial states:
	\begin{itemize}
		\item Product state: $|\Psi\rangle_1 = |\uparrow\downarrow\downarrow\cdots\rangle$
		\item Entangled pair: $|\Psi\rangle_{\mathrm{S}} = |S\rangle|\downarrow\downarrow\cdots\rangle$ with $|S\rangle = \frac{1}{\sqrt{2}} (|\uparrow\downarrow\rangle -|\downarrow\uparrow\rangle)$
		\item GHZ state: $|\Psi\rangle_{\mathrm{GHZ}} = \frac{1}{\sqrt{2}} (|\uparrow\uparrow\cdots\rangle + |\downarrow\downarrow\cdots\rangle)$
	\end{itemize}
	These represent common experimental preparations with varying entanglement structures, allowing us to probe state dependent fidelity behavior.
	
	In Sec.~\ref{sec:SWAP}, we analyze SWAP gate sequences, while Sec.~\ref{sec:generalU} extends to randomly generated general operations. Throughout, we assume perfect control operations (no noise in Eq.~\eqref{eq:H_ctrl}) to isolate connectivity noise effects.

	\section{Results}
	\subsection{Optimal gate duration of SWAP operations}\label{sec:SWAP}
	
	In quantum information processing, SWAP gates play a crucial role in enhancing connectivity by enabling quantum state transport between non-adjacent qubits. The operation time for a single SWAP gate is given by $\tau = \pi/(4J)$, where $J$ corresponds to the qubit-qubit coupling strength $J_{ij}$ in the control Hamiltonian (Eq.~\ref{eq:H_ctrl}).
	
	Using the characteristic coupling strength $J \approx 10$\,MHz from Table~\ref{tab:params}, the gate duration scales as $\tau \approx 80\,\text{ns} \times (100/J\lambda_0^{-1})$. Thus, our scanned range $J/\lambda_0 \in [1, 1000]$ corresponds to $\tau$ from $\sim 8\,\mu\text{s}$ down to $\sim 8\,\text{ns}$. The intermediate coupling regime ($J/\lambda_0 \sim 10$--$100$), where optimal points appear, corresponds to $\tau \sim 80$--$800\,\text{ns}$, which is longer than typical high-fidelity gate durations ($\sim 10$--$50\,\text{ns}$) but still within practical coherence windows.
	
\begin{figure}[t]
		\centering
		\includegraphics[width=0.8\linewidth]{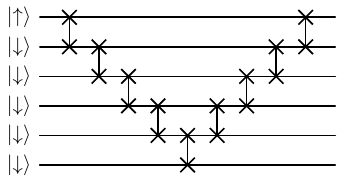}
		\caption{Quantum circuit implementing SWAP operations on product state $|\Psi\rangle_{1}$ for system size of $L = 6$ qubits as a representative example.
			The sequence transfers qubit states around the ring, with gate duration determined by the coupling strength $J$.}
		\label{fig:qc_swap100}
	\end{figure}

	We first investigate a quantum circuit composed of sequential SWAP operations, as illustrated in Fig.~\ref{fig:qc_swap100} for a representative system size of $L=6$. Starting from the unentangled product state $|\Psi\rangle_{1} = |\uparrow\downarrow\downarrow\cdots\rangle$, we perform a series of SWAP gates that transport the spin up qubit around the ring and back to its original position. The fidelity of this operation sequence is computed as the probability of obtaining the expected final state under ideal (noise free) evolution.
	
\begin{figure}[t]
		\centering
		\includegraphics[width=1\linewidth]{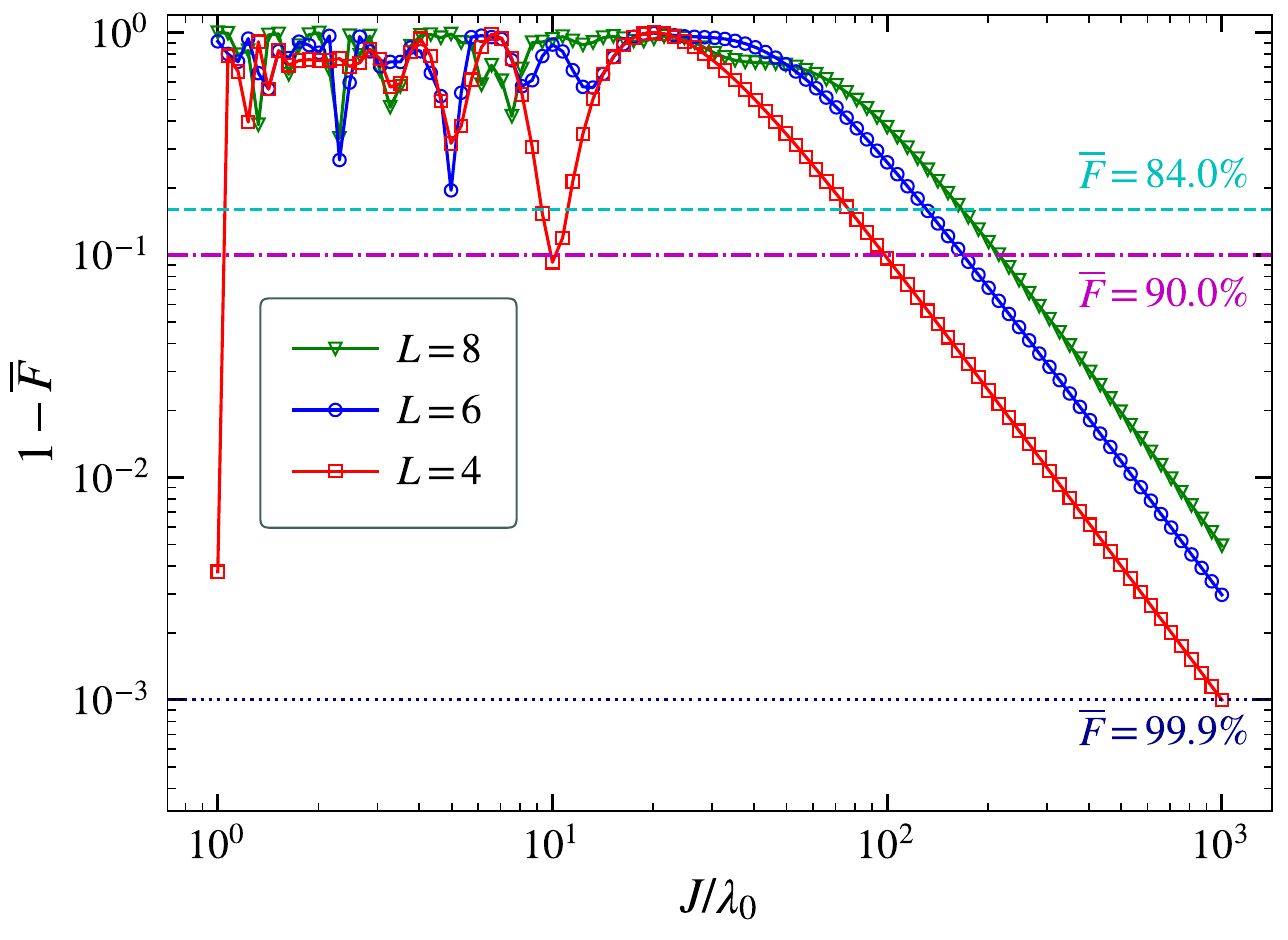}
		\caption{Average infidelity of SWAP operations versus $J/\lambda_0$ (spanning $10^{0}$ to $10^{3}$) for different system sizes $L$. For a characteristic coupling $J \approx 10$\,MHz, the corresponding gate duration $\tau$ ranges from $\sim 8\,\mu$s ($J/\lambda_0=1$) to $\sim 8\,\text{ns}$ ($J/\lambda_0=1000$).
			Horizontal dashed lines indicate fidelity benchmarks at 84\%, 90\%, and 99.9\%.
			Optimal operation points (minima) emerge in the intermediate coupling regime ($J/\lambda_0 < 100$), with the $L=4$ device achieving 90\% fidelity near $J/\lambda_0 = 10$.}
		\label{fig:dip}
	\end{figure}

	We systematically vary the ratio $J/\lambda_0$ from 1 to $10^3$, covering the weak to strong coupling regimes. Figure~\ref{fig:dip} shows the average infidelity as a function of $J/\lambda_0$ for different system sizes. For each $J/\lambda_0$ value, we sample 1,500 noise configurations from Gaussian distributions and compute the mean fidelity. In ideal conditions with minimal noise or strong pulses, fidelity typically improves monotonically with increasing $J/\lambda_0$, as demonstrated in previous studies~\cite{Foulk2022,Robert2020}. However, practical experimental constraints—where noise cannot be arbitrarily suppressed and pulse strengths face technical limits—make the intermediate coupling regime ($J/\lambda_0 < 10^2$) particularly relevant.
	
	Our results reveal a non-monotonic behavior in this experimentally significant regime. Rather than a steady improvement, the infidelity exhibits distinct minima—optimal operation points—where fidelity is maximized despite substantial noise levels. For the $L=4$ device, a fidelity of $\overline{F}=90\%$ (comparable to current experimental benchmarks~\cite{Kandel2019,Sigillito2019}) is achieved near $J/\lambda_0 = 10$. This demonstrates that performance typically expected under 1\% noise levels can be attained even with 10\% noise through optimal gate duration selection. Similar optimal points appear for larger system sizes, though the achievable fidelity decreases with increasing qubit number.
	
	\subsection{Optimal gate duration of general operations}\label{sec:generalU}
	
	We now extend our analysis to general quantum operations beyond simple SWAP gates. To characterize the complexity of these operations, we employ the distance to identity metric~\cite{NIELSEN2002}:
	\begin{equation}
		D = \overline{\left| \langle \Psi_0 | R |\Psi_0 \rangle \right|}^2 = \frac{d+|\tr R|^{2}}{d(d+1)},
	\end{equation}
	where $d=2^L$ is the Hilbert space dimension. For each distance value $D$, we randomly generate a unitary transformation $R$ with that specific distance to identity. The generation procedure is detailed in Appendix~\ref{sec:Implementation}.
	
	Figure~\ref{fig:qc_GU1} illustrates the evolution of the product state $|\Psi\rangle_{1}$ under a general random quantum circuit for $L=6$. For each operation $R$, we compute the fidelity by averaging over 1,500 noise configurations following the same methodology as in Section~\ref{sec:SWAP}.
	
	\begin{figure}[t]
		\centering
		\includegraphics[width=0.8\linewidth]{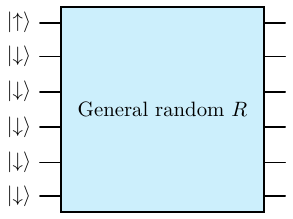}
		\caption{Evolution of product state $|\Psi\rangle_{1}$ under a general random quantum circuit $R$ for $L=6$ qubits.
			The left panel shows the initial state, while the right panel depicts the action of the randomly generated operation.}
		\label{fig:qc_GU1}
	\end{figure}
	
\begin{figure}[t]
		\centering
		\includegraphics[width=1\linewidth]{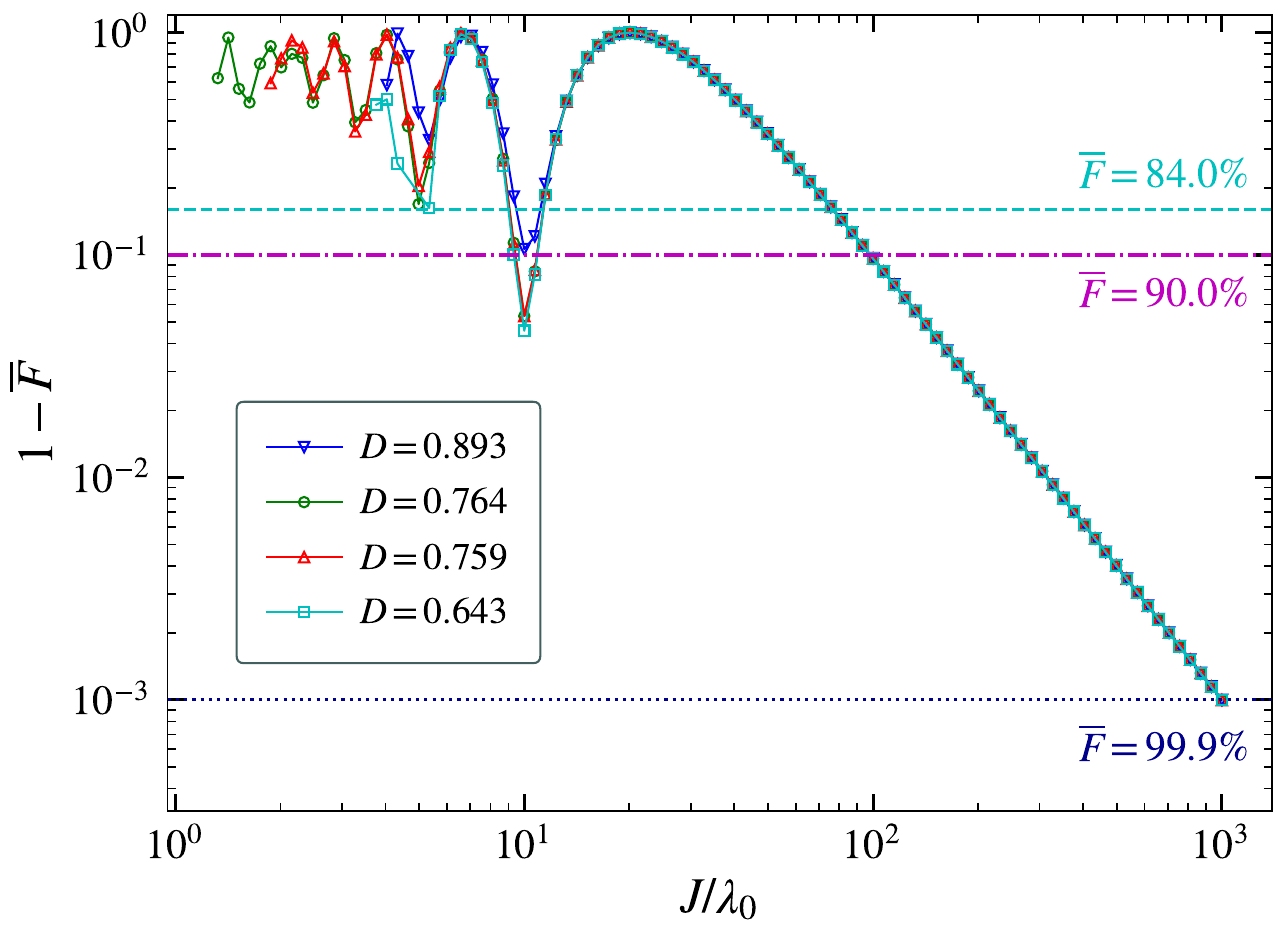}
		\caption{Fidelity of state $|\Psi\rangle_{1}$ for randomly generated quantum circuits with varying distances $D$ to identity for $L=4$ qubits. Each curve represents a single unitary operation $R$ with specific $D$, with fidelity averaged over 1,500 noise configurations. Optimal operation points occur at similar $J/\lambda_0$ values across different circuits.}
		\label{fig:qc_GU2}
	\end{figure}

	As shown in Fig.~\ref{fig:qc_GU2} (for $L=4$), optimal operation points emerge consistently across different randomly generated quantum circuits. Each curve corresponds to a specific unitary $R$ with a particular distance $D$, with fidelity values representing averages over noise configurations. We observe a clear correlation between the distance $D$ and the achievable fidelity: operations closer to the identity (smaller $D$) yield higher fidelity, as expected. More significantly, however, the optimal operation points for different quantum circuits all occur at approximately the same $J/\lambda_0$ value. This consistency indicates that once an optimal operating point is identified for one circuit in a given system, it can be leveraged to optimize other circuits without additional experimental overhead, enabling enhanced fidelity across multiple operations.
	
	\subsection{Impact of initial states on optimal points}\label{sec:states}
	
	We next investigate how different initial states affect operation fidelity, focusing on the entangled states $|\Psi\rangle_{\mathrm{S}}$ and $|\Psi\rangle_{\mathrm{GHZ}}$. Since the primary distinction among quantum circuits lies in fidelity magnitude rather than optimal point position, we restrict our analysis to SWAP gate sequences without loss of generality.
	
\begin{figure*}[t]
		\centering

		\includegraphics[width=1\linewidth]{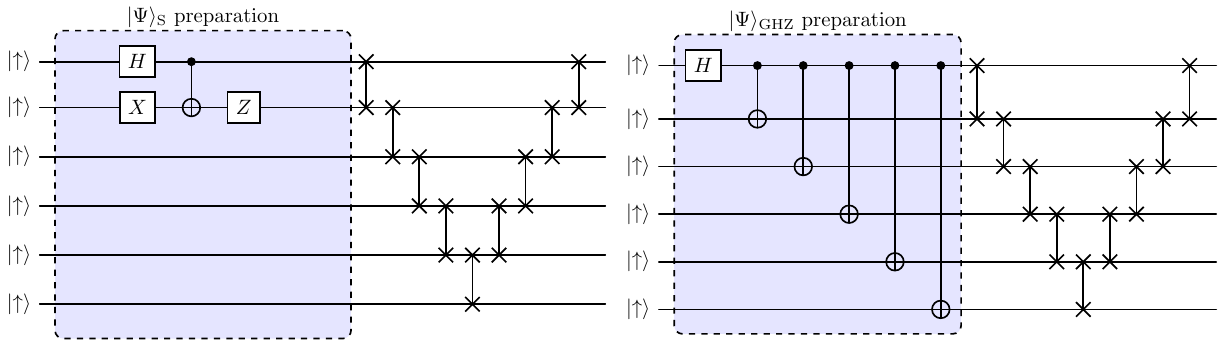}
		\put(-510, 130){\textbf{(a)}} 
		\put(-252, 130){\textbf{(b)}}    
		\caption{Quantum circuits for (a) $|\Psi\rangle_{\mathrm{S}}$ and (b) $|\Psi\rangle_{\mathrm{GHZ}}$ in a six-qubit system. Shaded regions indicate ideal state preparation, while subsequent SWAP operations are subject to connectivity noise.}
		\label{fig:circuitfor2}
	\end{figure*}

	Figure~\ref{fig:circuitfor2} illustrates the quantum circuits for preparing and evolving these entangled states in a six qubit system ($L=6$). For the singlet state $|\Psi\rangle_{\mathrm{S}}$, we sequentially swap the right qubit of the entangled pair around the ring and back, effectively distributing entanglement along the ring. For the GHZ state $|\Psi\rangle_{\mathrm{GHZ}}$, where all qubits are initially entangled, we perform consecutive swaps between adjacent qubits. The initial state preparation (shaded regions) is assumed ideal, enabling direct comparison of noise effects during the SWAP operations.
	
	%
	
\begin{figure}[t]
		\centering
		\includegraphics[width=1\linewidth]{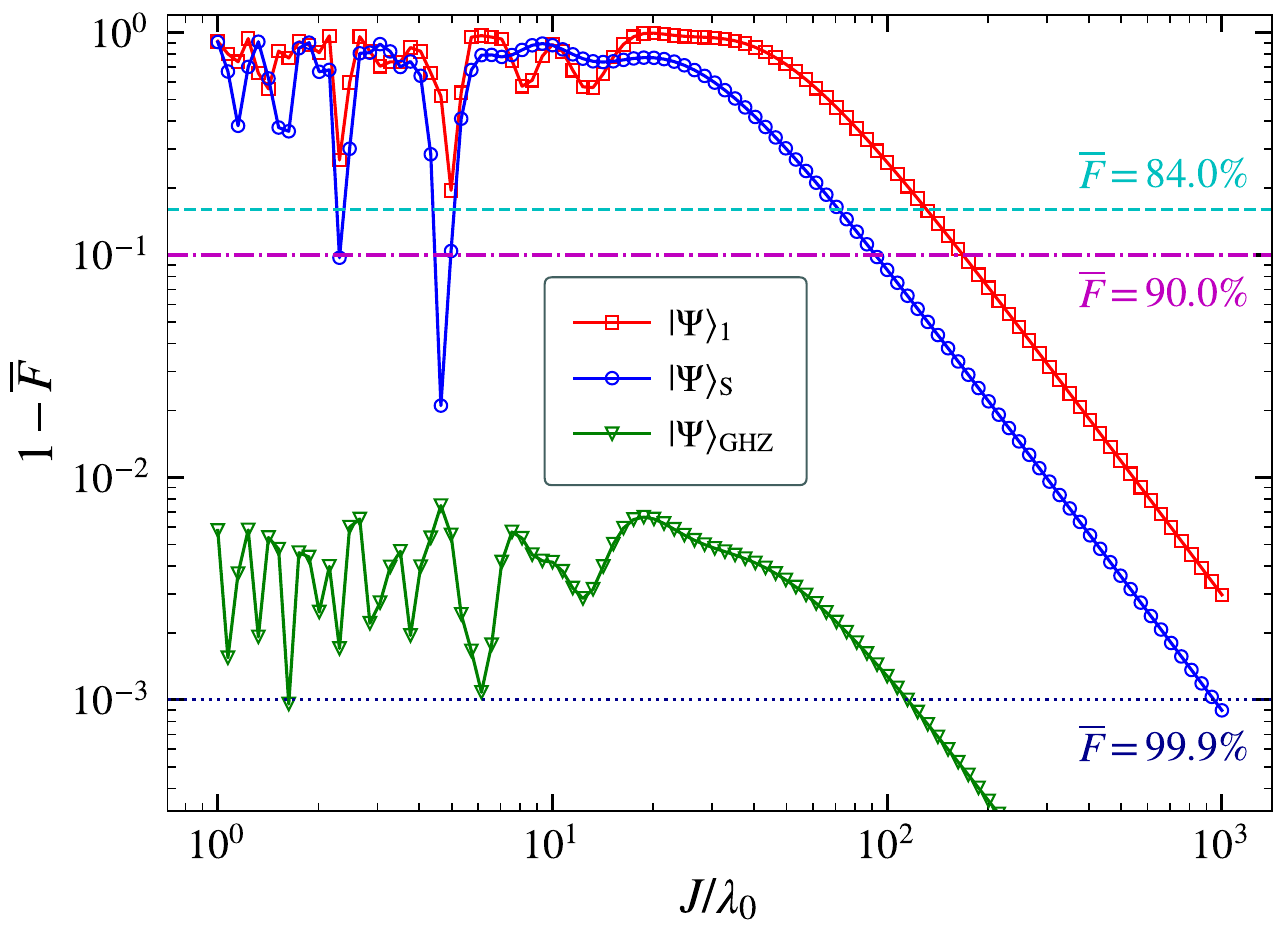}
		\caption{Fidelity of different initial states under SWAP operations for $L=6$. The fully entangled GHZ state achieves the highest fidelity, reaching 99.9\% at several optimal points, while the singlet state outperforms the product state.}
		\label{fig:state}
	\end{figure}

	As shown in Fig.~\ref{fig:state}, all three initial states exhibit clear optimal operation points under SWAP sequences. However, several important observations emerge. First, the positions of these optimal points vary among states, indicating that even with fixed device parameters and identical circuits, the optimal operating condition depends on the initial quantum state. More remarkably, contrary to the expectation that entangled states would suffer greater decoherence under repeated noisy operations, we observe the opposite trend: the fully entangled GHZ state achieves the highest overall fidelity, with several optimal points approaching the quantum error correction threshold of 99.9\%. The partially entangled singlet state $|\Psi\rangle_{\mathrm{S}}$ also outperforms the unentangled product state $|\Psi\rangle_{1}$.
	
	This suggests that higher initial entanglement can enhance robustness against connectivity noise in our system. The superior performance of the GHZ state may reflect favorable symmetry alignment between the initial state and the SWAP circuit structure, though the precise role of state-circuit symmetry requires further investigation~\cite{CABRERA2007,Sudha2011}. These results have practical implications for experimental implementations using transmon qubit rings, indicating that carefully chosen initial states can significantly improve operational fidelity.
	
	\subsection{Origin of the fidelity optimal operation point}\label{sec:mechanism}
	
	Our numerical results demonstrate that the quantum state fidelity in transmon qubit rings exhibits characteristic optimal operation points. This phenomenon has practical significance for real quantum circuits: under substantial noise or extended operation times, there exists an optimal evolution duration—equivalently, an optimal gate duration—that maximizes fidelity. From an experimental perspective, the position of this optimum is more critical than the exact fidelity value, as it remains largely independent of the specific circuit once system parameters are fixed.
	
	To elucidate the physical origin of this effect, we analyze an analytically solvable two qubit model with controlled approximations (see Appendix~\ref{app:twoqubit}). In this framework, the fidelity optimal point for the $|\uparrow\downarrow\rangle$ state arises from the intrinsic oscillatory nature of quantum coherence. Since $|\uparrow\downarrow\rangle$ is not an eigenstate of the noise Hamiltonian, its coherence exhibits temporal oscillations that periodically restore fidelity. In contrast, the GHZ state shows no such behavior in this simplified model due to the omission of superconducting qubit frequency terms, which effectively renders it an eigenstate of the Hamiltonian, thereby suppressing coherence oscillations. When realistic frequency terms are included, we expect GHZ type states to display analogous fidelity revivals and optimal point behavior.
	
	The universality of this mechanism stems from a general principle: any initial state that deviates from the Hamiltonian eigenbasis will exhibit oscillatory coherence dynamics under quasistatic noise. This characteristic is typical for systems dominated by slow noise fluctuations whose timescale exceeds that of the intrinsic quantum evolution~\cite{Wendin2017}. For other noise types—such as high frequency broadband (Markovian), low frequency narrowband (non-Markovian), or more complex forms including periodic, burst ike, or strongly correlated noise—the fidelity behavior becomes considerably richer. In these regimes, standard analyses based on simple coherence oscillations are inadequate, and approaches such as Lindblad master equations, hierarchical equations of motion (HEOM), Floquet theory, quantum trajectory methods, or many body simulations are required~\cite{Gorini1976,Becker2025,Zhang2024,Zhou2023,Carmichael1993,Dalibard1992}. Whether fidelity optimal points persist under such generalized noise conditions remains an open question for future investigation.

	\section{A machine learning approach to predict the optimal operation point}\label{sec:ML}
	
	\subsection{Machine learning framework}
	
	The numerical analyses presented thus far determine optimal operation points only for systems with fixed parameters. In experimental settings, however, both CPW coupling strengths and noise distributions can vary significantly and must often be characterized individually. Since these factors directly influence the optimal operation point—and consequently the ideal circuit duration—performing exhaustive simulations or measurements for every configuration would be computationally and experimentally prohibitive.
	
	To address this challenge, we employ a supervised machine learning approach based on neural networks to predict the optimal operation point for a given quantum device using only limited simulation data. Once trained, this model can estimate the optimal circuit duration for new parameter regimes defined by different CPW coupling strengths and noise characteristics.
	
\begin{figure}[t]
		\centering
		\includegraphics[width=1.0\linewidth]{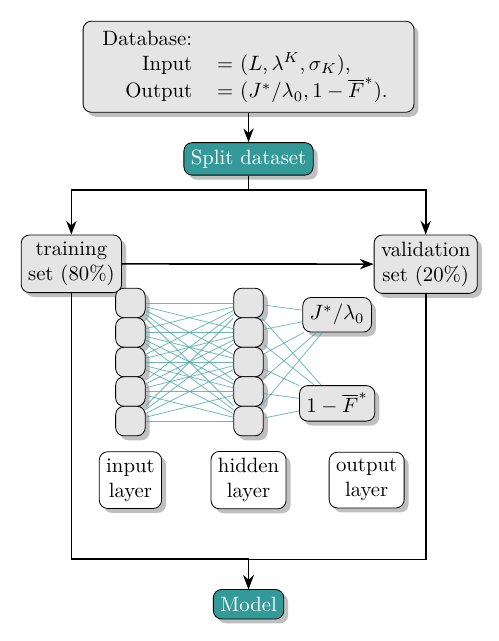}
		\caption{Supervised learning workflow using a multilayer perceptron neural network. Input parameters are system size $L$, CPW noise strength $\lambda^{K}$, and noise distribution width $\sigma_{K}$; outputs are the optimal operation point position and corresponding infidelity. The dataset is divided into training (80\%) and validation (20\%) subsets. For clarity, the diagram shows a simplified network with one hidden layer of six neurons; the actual model employs two hidden layers with ten neurons each, ReLU activations, and a linear output layer for two dimensional regression.}
		\label{fig:flow}
	\end{figure}

	The overall workflow, illustrated in Fig.~\ref{fig:flow}, proceeds as follows. We perform supervised learning by partitioning the available dataset into training and validation subsets, using labeled data to fit the model and then testing its predictive performance on unseen examples. Each data point is characterized by an input vector comprising the system size $L$, CPW noise strength $\lambda^{K}$, and noise distribution width $\sigma_{K}$, while the corresponding output contains the optimal operation point position $(J^{*}/\lambda_{0})$ and its associated infidelity $(1-\overline{F}^{*})$.
	
	As the cost function, we adopt the mean squared error (MSE) loss~\cite{Kevin2012,Kevin2022,Heaton2018}:
	\begin{equation}
		\text{MSE} = \frac{1}{n} \sum_{i=1}^{n} (y_i - \hat{y}_i)^2
	\end{equation}
	where $n$ denotes the total number of data points, $y_i$ the observed value, and $\hat{y}_i$ the model prediction. We further evaluate performance using the coefficient of determination:
	\begin{equation}
		R^2 =  1 - \frac{\sum_{i=1}^{n} (y_i - \hat{y}_i)^2}{\sum_{i=1}^{n} (y_i - \bar{y})^2}
	\end{equation}
	where $\bar{y}$ denotes the mean of observed values. In our case, each $y_i$ corresponds to a two-dimensional vector $(J^{*}/\lambda_0, 1-\overline{F}^{*})$. These complementary metrics provide a consistent framework for evaluating model performance.
	
	We employ a multilayer perceptron (MLP) architecture with two hidden layers of ten neurons each, using Rectified Linear Unit (ReLU) activations for nonlinear expressivity~\cite{Sudjianto2020}. The output layer uses linear activation appropriate for regression tasks. The model is trained using the Adam optimizer~\cite{Kingma2017}, which combines advantages of Adagrad and RMSProp for efficient optimization~\cite{Duchi2011,Ruder2016,Wilson2017}.
	
	\subsection{Data extraction and neural network training}
	
	To construct the training dataset, we compute fidelity as a function of $J/\lambda_0$ for system sizes $L = 4, 6,$ and 8, varying CPW noise strength within $\lambda^{K}/\lambda_{0} \in [3,20]$ and noise-distribution width within $\sigma_{K} \in [0.01,0.1] \times \lambda^{K}$. From these simulations, we extract optimal operation point positions and corresponding fidelities to form the target outputs.
	
\begin{figure}[tb]
		\centering
		\includegraphics[width=1.\linewidth]{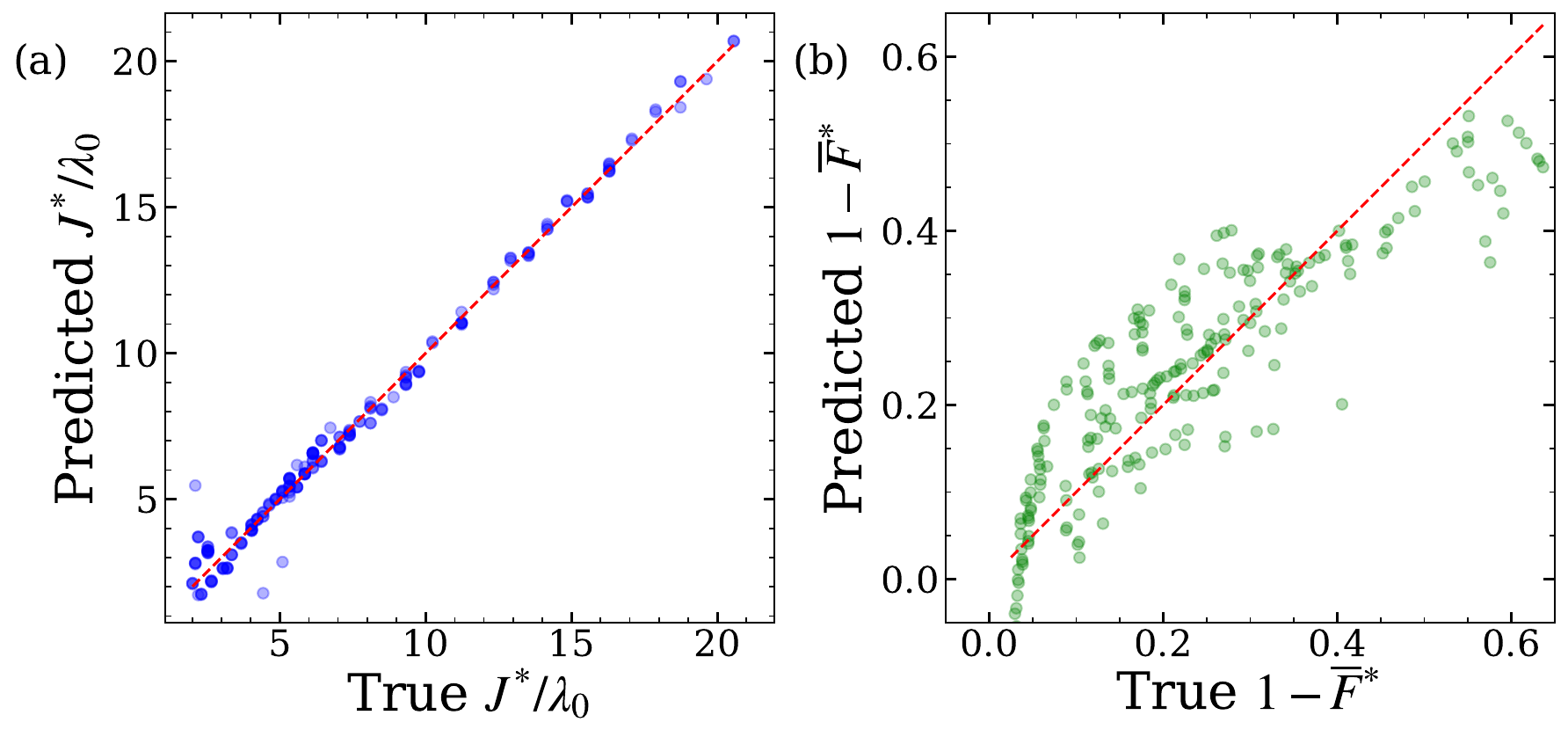}
		\caption{Comparison between predicted and true values for (a) optimal operation point position $J^{*}/\lambda_{0}$ and (b) infidelity $1-\overline{F}^{*}$. Position prediction shows excellent agreement (${\rm MSE}=0.31$, $R^{2}\approx0.99$), while fidelity prediction is less accurate (${\rm MSE}=0.0048$, $R^{2}\approx0.81$).}
		\label{fig:ML}
	\end{figure}

	Figure~\ref{fig:ML} shows the resulting predictions. The optimal point position prediction shows excellent agreement with validation data (${\rm MSE}=0.31$, $R^{2}\approx0.99$), while fidelity prediction is less accurate (${\rm MSE}=0.0048$, $R^{2}\approx0.81$). This difference arises because optimal point positions are more reliably determined from simulations and exhibit greater consistency across different circuits, as discussed in Section~\ref{sec:generalU}.
	
\begin{figure}[bt]
		\centering
		\includegraphics[width=1.\linewidth]{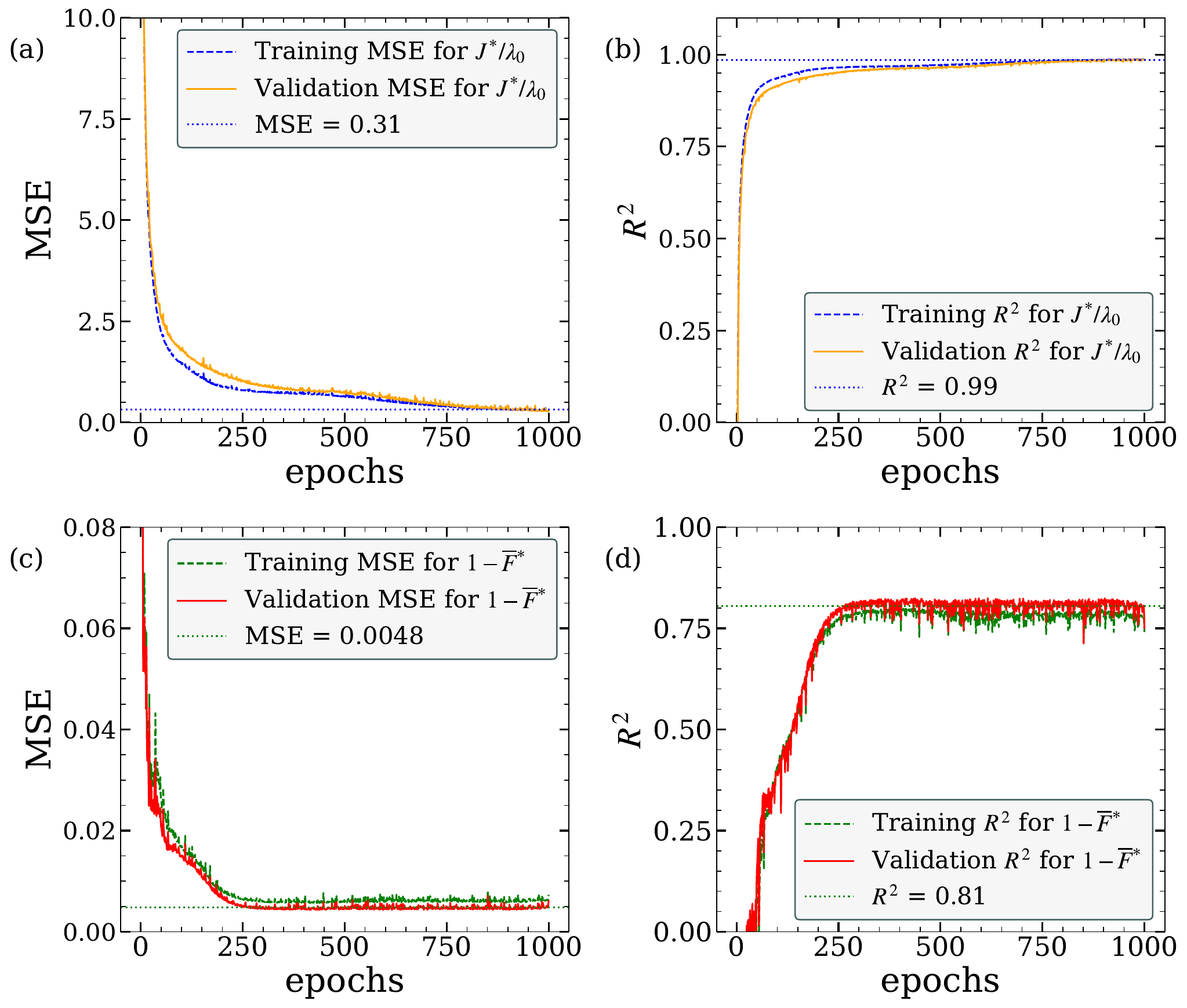}
		\caption{Evolution of MSE and $R^{2}$ during training for (a,b) optimal point position $J^{*}/\lambda_{0}$ and (c,d) infidelity $1-\overline{F}^{*}$. Both metrics converge after approximately 250 epochs.}
		\label{fig:MSE}
	\end{figure}

	Figure~\ref{fig:MSE} shows the evolution of training performance. Both MSE and $R^2$ metrics converge after approximately 250 epochs, confirming the stability and effectiveness of the neural network model. The accurate prediction of optimal operation point positions is particularly valuable experimentally, as it enables determination of ideal circuit durations without extensive characterization measurements.
	
	\section{Conclusions and discussions}\label{sec:conclusion}
	
	In summary, our investigation of transmon qubit systems reveals that optimal operation points emerge consistently across various quantum operations, even under substantial connectivity noise. The identification of these fidelity maxima in the intermediate coupling regime ($10 \lesssim J/\lambda_0 \lesssim 100$) provides a practical pathway to enhance gate performance in multiqubit devices without requiring unrealistic noise suppression. This finding is particularly relevant for implementing SWAP gates and general quantum operations that form the foundation of scalable quantum computing architectures.
	
	Our analysis demonstrates that the position of optimal operation points remains largely independent of the specific quantum circuit, enabling efficient optimization across diverse operations. Moreover, the significant dependence of fidelity on initial states reveals that entanglement structure and symmetry alignment play crucial roles in noise resilience. The exceptional performance of GHZ states under SWAP operations, reaching fidelities compatible with quantum error correction thresholds, highlights the importance of state-circuit symmetry matching for robust quantum information processing. The underlying mechanism, rooted in coherence oscillations under quasistatic noise, provides a universal framework for understanding fidelity dynamics in superconducting qubit systems.
	
	The integration of machine learning techniques further enhances the practical utility of our findings by enabling efficient prediction of optimal operation points across varying device parameters. Our supervised neural network approach accurately identifies optimal circuit durations without exhaustive simulations, adapting effectively to variations in CPW coupling strengths and noise distributions. This data driven strategy significantly reduces the experimental overhead for optimizing quantum operations in diverse device configurations.
	
	Overall, our results provide valuable insights for designing next generation quantum processors that leverage optimal gate duration tuning to achieve high fidelity operations under realistic noise conditions. By identifying and exploiting the inherent trade-offs between circuit depth and noise accumulation, these strategies offer a pragmatic approach to enhancing quantum computational performance in the noisy intermediate scale quantum era.
	
	\acknowledgments
	The authors thank Liu Jie, Wu Jiaohao, He Minquan, Zhu Yan and Ni Xiaotong for fruitful discussions. This work is supported by the National Natural Science Foundation of China (Grant No. 12474489), the Shenzhen Fundamental Research Program (Grant No. JCYJ20240813153139050), the Guangdong Provincial Quantum Science Strategic Initiative (Grant No. GDZX2203001, GDZX2403001), and Quantum Science and Technology-National Science and Technology Major Project (Grant No. 2021ZD0302300).
	\section*{Declarations}
	The authors declare no conflict of interest.

	\appendix
	
	\section{Comprehensive Model for Transmon Qubits with Hybrid Connectivity}
	\label{sec:full_model}
	We present a comprehensive Hamiltonian model for systems of $L = 4$,6, and 8 transmon qubits arranged in a ring geometry with hybrid cavity-CPW connectivity, as illustrated in Fig.~\ref{fig:connectivity}~\cite{Koch2007}.
	The model includes both static and dynamic noise contributions that persist independently of the applied control operations, indicating that the noise characteristics are determined primarily by intrinsic device parameters and fabrication quality~\cite{Zhang2019,Zhao2025v2}.
	\subsection{Control Hamiltonian}
	Each transmon qubit is governed by the circuit Hamiltonian:\begin{equation}
		H_{0,i} = 4E_C^i (\hat{n}_i - n_g^i)^2 - E_J^i \cos \hat{\phi}_i,
	\end{equation}
	where $\hat{n}_i = -i\partial_{\phi_i}$ is the charge operator conjugate to phase $\hat{\phi}_i$, $E_C^i = e^2/(2C_{\Sigma}^i)$ is the charging energy with $C_{\Sigma}^i$ being the total capacitance, and $n_g^i$ is the offset charge. In the transmon regime ($E_J^i / E_C^i \gg 1$), we expand to fourth order:
	\begin{align}
		\cos \hat{\phi}_i &\approx 1 - \frac{1}{2}\hat{\phi}_i^2 + \frac{1}{24}\hat{\phi}_i^4 \\
		H_{0,i} &\approx \underbrace{\left[4E_C^i \hat{n}_i^2 + \frac{E_J^i}{2}\hat{\phi}_i^2\right]}_{\text{harmonic part}} - \underbrace{\frac{E_J^i}{24}\hat{\phi}_i^4}_{\text{anharmonicity}} - 8E_C^i n_g^i \hat{n}_i
	\end{align}
	Quantization with ladder operators yields:
	\begin{align}
		\hat{\phi}_i &= \varphi_i (\hat{b}_i^\dagger + \hat{b}_i), \quad
		\varphi_i = \left( \frac{2E_C^i}{E_J^i} \right)^{1/4} \\
		\hat{n}_i &= i n_i (\hat{b}_i^\dagger - \hat{b}_i), \quad
		n_i = \frac{1}{2} \left( \frac{E_J^i}{8E_C^i} \right)^{1/4}
	\end{align}
	resulting in the Duffing oscillator Hamiltonian:
	\begin{equation}
		H_{0,i} = \hbar\omega_i^q \hat{b}_i^\dagger \hat{b}_i + \frac{\alpha_i}{2} \hat{b}_i^\dagger \hat{b}_i^\dagger \hat{b}_i \hat{b}_i
	\end{equation}
	where $\hbar\omega_i^q = \sqrt{8E_C^i E_J^i} - E_C^i$ is the qubit frequency and $\alpha_i = -E_C^i$ is the anharmonicity.
	
	In the hybrid architecture, neighboring qubits interact via shared superconducting cavities.
	The bare cavity-qubit interaction Hamiltonian is given by
	\begin{equation}
		H_{\text{int}}^{\text{cav}} = \sum_{i,k} g_{ik} (\hat{b}_i^\dagger a_k + \hat{b}_i a_k^\dagger),
	\end{equation}
	where $a_k$ ($a_k^\dagger$) denotes the annihilation (creation) operator of cavity mode $k$, and $g_{ik}$ is the coupling strength between qubit $i$ and cavity $k$.
	The detuning between the qubit and cavity frequencies is defined as $\Delta{ik} = \omega_i^{q} - \omega_k^{\text{cav}}$.
	In the dispersive regime, where $|\Delta_{ik}| \gg |g_{ik}|$, we apply the rotating-wave approximation (RWA) and perform a Schrieffer-Wolff transformation
	\begin{equation}
		U = \exp\left[ \sum_{i,k} \frac{g_{ik}}{\Delta_{ik}} (\hat{b}_i^\dagger a_k - \hat{b}_i a_k^\dagger) \right],
	\end{equation}
	yielding the effective qubit-qubit coupling:
	\begin{equation}
		H_{\text{coup}}^{\text{cav}} = \sum_{\langle i,j\rangle} 4J_{ij} (\hat{b}_i^\dagger \hat{b}_j + \text{h.c.}),
	\end{equation}
	with coupling strength
	\begin{equation}
		J_{ij} = \frac{g_{ik}g_{jk}}{8} \left( \frac{1}{\Delta_{ik}} + \frac{1}{\Delta_{jk}} \right).
	\end{equation}
	We note that the coupling strength $J_{ij}$ depends on the detunings $\Delta_{ik}$ and $\Delta_{jk}$. In experimental systems, tuning the frequency of a qubit (thus changing its detuning from a resonator) will affect all coupling terms involving that qubit. For instance, varying $\Delta_{ik}$ to adjust $J_{ij}$ will also modify $J_{i\ell}$ for other qubits $\ell$ coupled to the same resonator $k$. This interdependence implies that in practice, calibrating multiple couplings may require global optimization over several parameters. In our theoretical model, we treat the $J_{ij}(t)$ as independently tunable parameters, which is an idealization that allows us to focus on the generic effects of connectivity noise. Advanced control techniques, such as using dedicated tunable couplers, can provide a high degree of independent control over effective couplings, approximating this idealization. Similar considerations apply to the CPW-mediated couplings $K_{ij}$.
	
	Mapping to the computational subspace $\{|g\rangle, |e\rangle\}$:
	\begin{align}
		\hat{b}_i^\dagger &\to \sigma_i^+ = \frac{\sigma_i^x + i\sigma_i^y}{2} \\
		\hat{b}_i &\to \sigma_i^- = \frac{\sigma_i^x - i\sigma_i^y}{2}
	\end{align}
	gives the ideal cavity-mediated interaction:
	\begin{equation}
		H_{\text{coupling}}^{\text{ideal}} = \sum_{\langle i,j\rangle} g_{ij}^{\text{cav}}(t) (\sigma_i^x \sigma_j^x + \sigma_i^y \sigma_j^y)
		\label{eq:Hcav_ideal}
	\end{equation}
	where $g_{ij}^{\text{cav}}(t) = 2J_{ij}$.
	This tunable interaction represents the intended qubit-qubit coupling mediated by the cavity mode.
	
	In an analogous manner, we derive the Hamiltonian for the CPW coupling by considering the CPW-transmon interaction:
	\begin{equation}
		H_{\text{int}}^{\text{cpw}} = \sum_{i,k} g_{ik}^{\text{cpw}}\hat{n}i \left(b_k^\dagger + b_k\right),
	\end{equation}
	where $b_k$ ($b_k^\dagger$) denotes the annihilation (creation) operator of CPW mode $k$, and $g_{ik}^{\text{cpw}}$ is the coupling strength between qubit $i$ and mode $k$.
	
	Virtual photon exchange through the CPW mediates an effective qubit-qubit interaction described by
	\begin{equation}
		H_{\text{eff}}^{\text{cpw}} = \sum_{i,j,k} \frac{g_{ik}^{\text{cpw}} g_{jk}^{\text{cpw}}}{\omega_k^{\text{cpw}}}\hat{n}_i \hat{n}_j .
	\end{equation}
	After applying RWA and mapping to the qubit basis, we obtain the effective CPW coupling Hamiltonian:
	\begin{equation}
		H_{\text{coupling}}^{\text{cpw}} = \sum_{\{i,j\}} g_{ij}^{\text{cpw}}\left(\sigma_i^+ \sigma_j^- + \text{h.c.}\right),
	\end{equation}
	where $\{i,j\}$ labels the non-adjacent qubits, and
	\begin{equation}
		g_{ij}^{\text{cpw}} = \sum_k \frac{g_{ik}^{\text{cpw}} g_{jk}^{\text{cpw}}}{\omega_k^{\text{cpw}}}n_i n_j .
	\end{equation}

	Single-qubit control is achieved through a combination of microwave driving and flux biasing:
	\begin{itemize}
		\item \textbf{X/Y control}:
		Microwave pulses are applied via individual transmission lines to induce rotations in the $x$-$y$ plane of the Bloch sphere.
		\item \textbf{Z control}:
		Flux biasing is used to tune the qubit transition frequency, enabling dynamic $z$-axis rotations and frequency detuning.
	\end{itemize}
	The corresponding microwave-drive Hamiltonian for qubit $i$ is given by
	\begin{equation}
		H_{\mathrm{drive}}^i = \mathcal{E}_i(t) \cos(\omega_d^i t + \phi_i(t)) \hat{n}_i
	\end{equation}
	where $\mathcal{E}_i(t)$ is the drive amplitude, $\omega_d^{i}$ the drive frequency, $\phi_i(t)$ the time-dependent phase, and $\hat{n}_i$ the charge operator.
	
	Applying the  RWA near resonance ($\omega_d^i \approx \omega_q^i$), the drive Hamiltonian becomes:
	\begin{align}
		H_{\mathrm{drive}}^i &\approx \frac{\mathcal{E}_i(t)}{2} \left[ e^{i(\omega_d^i t + \phi_i(t))} \sigma_i^- + \mathrm{h.c.} \right] \nonumber \\
		&= \frac{\Omega_i^x(t)}{2} \left( \cos\phi_i(t) \sigma_i^x + \sin\phi_i(t) \sigma_i^y \right)
	\end{align}
	where $\Omega_i^x(t) = |\mathcal{E}_i(t) \kappa_i|$ denotes the effective Rabi rate.
	
	The flux-mediated frequency shift~\cite{Koch2007} is expressed as
	\begin{equation}
		\Delta_i(t) = \frac{\partial \omega_q^i}{\partial \Phi} \delta\Phi_i(t)
	\end{equation}
	where $\delta\Phi_i(t)$ is the applied flux bias.
	This leads to the $Z$-control Hamiltonian
	\begin{equation}
		H_{\mathrm{zshift}}^i = \frac{\Delta_i(t)}{2} \sigma_i^z,
	\end{equation}
	which generates unitary rotations of the form $e^{-i\theta \sigma_z/2}$ on the Bloch sphere.
	
	Combining all contributions, the complete control Hamiltonian is
	\begin{equation}
		\begin{aligned}
			H_{\mathrm{ctrl}} &= \underbrace{\sum_i \left[ \Omega_i^x(t) \left( \cos\phi_i(t) \sigma_i^x + \sin\phi_i(t) \sigma_i^y \right) + \frac{\Delta_i(t)}{2} \sigma_i^z \right]}_{\text{Single-qubit control}} \\
			&+ \underbrace{\sum_{\langle i,j\rangle} 2J_{ij}(t) \left( \sigma_i^+ \sigma_j^- + \mathrm{h.c.} \right)}_{\text{Cavity couplings}} \\
			&+ \underbrace{\sum_{\{i,j\}} 2K_{ij}(t) \left( \sigma_i^+ \sigma_j^- + \mathrm{h.c.} \right)}_{\text{CPW couplings}}
		\end{aligned}
	\end{equation}
	where $\langle i,j\rangle$ labels adjacent qubits, $\{i,j\}$ labels non-adjacent qubits, and the effective coupling strengths are defined as $J_{ij}(t) = g_{ij}^{\mathrm{cav}}(t)/2$ and $K_{ij} = g_{ij}^{\mathrm{cpw}}/2$.

	\subsection{Always-on noise Hamiltonian}
	\label{ssec:noise}
	\subsubsection{Parasitic capacitance-induced noise}
	Direct capacitive coupling between qubits introduces an additional always-on interaction described by
	\begin{equation}
		H_{\text{cap}} = \hbar g_c \hat{n}_i \hat{n}_j, \quad g_c = \frac{2e^2}{\hbar} \frac{C_g}{\sqrt{C_{\Sigma}^i C_{\Sigma}^j}},
	\end{equation}
	where $C_g $ is the mutual capacitance. Following the standard operator mapping convention~\cite{Kong2018thesis},
	\begin{align}
		\hat{n}_i &= n_i \sigma_i^y, \quad 
		\sigma_i^y = i(\sigma_i^- - \sigma_i^+) \\
		n_i &= \frac{1}{2} \left( \frac{E_J^i}{8E_C^i} \right)^{1/4} .
	\end{align}
	Substituting and truncating to the qubit subspace yields
	\begin{align}
		H_{\text{cap}} &= \hbar g_c (n_i \sigma_i^y)(n_j \sigma_j^y) \nonumber \\
		&= \hbar g_c n_i n_j \sigma_i^y \sigma_j^y.
	\end{align}
	The positive sign reflects the repulsive nature of the underlying charge-charge interaction.
	
	Under ideal conditions, $C_g$ corresponds only to the designed coupling capacitance.
	In practice, fabrication imperfections introduce parasitic capacitance $C_g^{\text{para}}$, leading to a modified effective coupling strength,
	\begin{equation}
		g_c \to g_c' = \frac{2e^2}{\hbar} \frac{C_g + C_g^{\text{para}}}{\sqrt{C_{\Sigma}^i C_{\Sigma}^j}} .
	\end{equation}
	
	This modification gives rise to a quasi-static noise term,
	\begin{equation}
		H_{\text{noise}}^{yy} = \sum_{\langle i,j\rangle} \lambda_{ij}^{J} \sigma_i^y \sigma_j^y,
		\label{eq:Hnoise_yy}
	\end{equation}
	with noise strength
	\begin{equation}
		\lambda_{ij}^{J} = \hbar g_c' n_i n_j = \frac{\sqrt{E_C^i E_C^j}}{\hbar} \cdot \frac{\beta_{ij}}{\sqrt{\omega_i \omega_j}} > 0
	\end{equation}
	where $\beta_{ij} = 2 C_g^{\text{para}} / C_{\Sigma}^{ij}$ quantifies the fractional parasitic capacitance.
	This always-on noise interaction remains active even when the cavity-mediated coupling is nominally turned off.

	\subsubsection{CPW photon loss-induced noise}
	\label{ssec:cpw_noise}
	The quasi-static noise in CPW-mediated couplings originates from photon loss within the transmission line. We derive this noise term starting from the fundamental light-matter interaction and show how photon dissipation induces stochastic fluctuations in the effective qubit-qubit couplings. Each CPW segment supports a continuum of photon modes coupled to transmon qubits via dipole interactions. The corresponding interaction Hamiltonian is
	\begin{equation}
		H_{\mathrm{int}}^{\mathrm{cpw}} = \sum_{i,k} \int d\omega \, g_k^i(\omega) \left( a_k^\dagger(\omega) \sigma_i^- + a_k(\omega) \sigma_i^+ \right),
	\end{equation}
	where $a_k(\omega)$ ($a_k^\dagger(\omega)$) are photon annihilation (creation) operators for mode $\omega$ in CPW segment $k$, and $g_k^i(\omega)$ idenotes the frequency-dependent coupling strength.
	
	Under the Markovian approximation, the ideal effective qubit-qubit exchange coupling for non-adjacent pairs $\{i,j\}$ takes the form:
	\begin{equation}
		H_{\mathrm{coup}}^{\mathrm{ideal}} = \sum_{\{i,j\}} 2K_{ij} (\sigma_i^+ \sigma_j^- + \mathrm{h.c.}).
	\end{equation}

	Photon loss occurs at a rate $\kappa = \bar{\omega}/Q$, where $Q \sim 10^4$-$10^5$ is the CPW quality factor.
	Incorporating dissipation using the input-output formalism, the Heisenberg equation for the photon field acquires a Langevin noise term:
	\begin{equation}
		\dot{a}_k(\omega) = -\frac{\kappa}{2} a_k(\omega) - i \sum_i g_k^i(\omega) \sigma_i^- + \sqrt{\kappa} a_{\mathrm{in},k}(\omega),
	\end{equation}
	where $a_{\mathrm{in},k}(\omega)$ represents the vacuum noise operators, satisfying:
	\begin{align}
		\langle a_{\mathrm{in},k}(\omega) \rangle &= 0 \\
		\langle a_{\mathrm{in},k}^\dagger(\omega) a_{\mathrm{in},k'}(\omega') \rangle &= \delta_{kk'} \delta(\omega - \omega') n_{\mathrm{th}}(\omega) \\
		\langle a_{\mathrm{in},k}(\omega) a_{\mathrm{in},k'}^\dagger(\omega') \rangle &= \delta_{kk'} \delta(\omega - \omega') [n_{\mathrm{th}}(\omega) + 1].
	\end{align}
	At zero temperature ($n_{\mathrm{th}} = 0$), these vacuum fluctuations generate stochastic variations in the photon field.
	
	Adiabatically eliminating the photon field yields an effective qubit-qubit interaction containing stochastic corrections.
	The time-dependent coupling amplitude is
	\begin{align}
		\lambda_{ij}^{K}(t) &= -\sum_k \iint d\omega d\omega' \, \frac{g_k^i(\omega) g_k^j(\omega')}{\Delta_{ik} \Delta_{jk}} \\
		&\times \left[ a_{\mathrm{in},k}(\omega) e^{-i\omega t} + a_{\mathrm{in},k}^\dagger(\omega') e^{i\omega' t} \right],
		\label{eq:noise_kernel}
	\end{align}
	where $\Delta_{ik} = \omega - \omega_i^q$. The photon loss rate $\kappa$ determines the correlation function of this stochastic coupling:
	\begin{equation}
		\langle \lambda_{ij}^{K}(t) \lambda_{mn}^{K}(t') \rangle \propto e^{-\kappa|t-t'|/2} \cos[\bar{\omega}(t-t')],
	\end{equation}
	indicating that higher photon loss ($\uparrow\kappa$) enhances temporal correlations on timescales shorter than $1/\kappa$.
	
	For gate operations slower than $1/\kappa$ but faster than overall system evolution, the noise can be regarded as quasi-static with Gaussian statistics:
	\begin{equation}
		\lambda_{ij}^{K}(t) \sim \mathcal{N}(\lambda_{ij,0}^{K}, \sigma_{K}^2)
	\end{equation}
	The mean coupling $\lambda_{ij,0}^{K}$ arises from the power of vacuum fluctuations,
	\begin{equation}
		\lambda_{ij,0}^{K} = \frac{\bar{g}_i \bar{g}_j \sqrt{\kappa}}{\sqrt{2\pi \bar{\omega}}} 
	\end{equation}
	where $\bar{g}_i = \sqrt{N \langle g_k^2 \rangle}$ represents the average coupling over $N \sim 10$ photon modes within the CPW bandwidth, and $\bar{\omega}$ is the mean mode frequency.
	
	The scaling $\lambda_{ij,0}^{K} \propto \sqrt{\kappa}$ highlights the fundamental link between photon loss and noise amplitude: increased photon dissipation amplifies vacuum fluctuations, thereby elevating the noise floor.
	Similarly, the variance $\sigma_{K}^2$ follows a comparable dependence on photon loss and mode density, $\sigma_K\propto \frac{\bar{g}_i \bar{g}_j \kappa^{3/4}}{\bar{\omega}^{1/2}}$, confirming the intrinsic connection between CPW photon loss ($\kappa = \bar{\omega}/Q$) and the statistical properties of coupling noise.
	
	The effective noise amplitude $\lambda_{ij}^{K}(t)$ scales as $\propto 1/(\Delta_{ik} \Delta_{jk})$. In practical devices, tuning the qubit frequency (thus changing $\Delta_{ik}$) to adjust the intended coupling $J_{ij}$ would simultaneously modify the noise statistics. In our theoretical model, however, we treat the noise amplitudes $\lambda_{ij}^{J}$ and $\lambda_{ij}^{K}$ as fixed parameters determined by static device properties. This simplification allows us to focus on the generic dependence of fidelity on the dimensionless ratio $J/\lambda_0$. We emphasize that the existence of optimal operation points is a robust phenomenon that persists even when noise amplitudes vary with detuning, as the underlying physical mechanism is governed by the ratio between coupling and noise energy scales. For experimental calibration, a global optimization over both coupling and noise parameters may be required.
	
	\subsection{Complete system Hamiltonian}
	\label{ssec:full_hamiltonian}
	
	The complete time-dependent Hamiltonian integrates all fundamental components of the quantum processor, including control dynamics and intrinsic noise mechanisms:
	\begin{align}
		H(t) = & \sum_i \Bigg[ \Omega_i^x(t) \left( \cos\phi_i(t) \sigma_i^x + \sin\phi_i(t) \sigma_i^y \right) \nonumber\\
		&+ \frac{\Delta_i(t)}{2} \sigma_i^z + \delta\omega_{i}(t)\sigma_i^z \Bigg] \nonumber \\
		& + \sum_{\langle i,j\rangle} \left[ 2J_{ij}(t) \left( \sigma_i^+ \sigma_j^- + \mathrm{h.c.} \right) + \lambda^{J}_{ij}(t) \sigma_i^y \sigma_j^y \right] \nonumber \\
		& + \sum_{\{i,j\}} \left[ 2K_{ij} +  \lambda^{K}_{ij}(t) \right] \left( \sigma_i^+ \sigma_j^- + \mathrm{h.c.} \right).
		\label{eq:H_full}
	\end{align}
	
	The corresponding parameter ranges used in simulations are summarized in Table~\ref{tab:params}.
	
	\begin{table}[h]
		\centering
		\caption{Simulation parameters for transmon devices, based on typical experimental ranges~\cite{Kong2018thesis,Wang2022}. The chosen values allow isolation of connectivity-noise effects; the results exhibit scaling invariance as discussed in the text.}
		\label{tab:params}
		\begin{tabular}{@{}lll@{}}
			\toprule
			& Parameter & Characteristic Value \\
			\hline
			\midrule
			\multicolumn{3}{l}{\textit{Single-Qubit Control Parameters}} \\
			\hline
			& $X$-drive amplitude, $\Omega_i^x(t)$ & 100 MHz \\
			& Dynamic detuning, $|\Delta_i(t)|$ & 20 MHz \\
			\hline
			\midrule
			\multicolumn{3}{l}{\textit{Single-Qubit Noise Parameters}} \\
			\hline
			& Frequency shift, $\delta\omega_i(t)$ & $1\times10^{-4}$--$8.5\times10^{-3}$ MHz \\
			& & (corresponds to $T_2^* \gtrsim 20\ \mu$s) \\
			\hline
			\midrule
			\multicolumn{3}{l}{\textit{Two-Qubit Coupling Parameters}} \\
			\hline
			& Cavity coupling strength, $|2J_{ij}(t)|$ & 20 MHz \\
			& CPW coupling strength, $|2K_{ij}|$ & 100 MHz \\
			\hline
			\midrule
			\multicolumn{3}{l}{\textit{Two-Qubit Noise Parameters}} \\
			\hline
			& Parasitic capacitance noise, $\lambda^{J}_{ij}(t)$ & 0.05--0.5 MHz \\
			& Photon loss noise, $\lambda^{K}_{ij}(t)$ & 1.0--5.0 MHz \\
			& Noise ratio, $\mathcal{R} = \lambda^{K}_{ij}/\lambda^{J}_{ij}$ & 3--20 \\
			\hline
			\bottomrule
		\end{tabular}
	\end{table}
	
	This unified model provides a physically consistent framework for simulating transmon-based quantum processors with $L=4$, $6$, and $8$ qubits.
	In this formulation, all noise terms arise naturally from device-level physics while remaining strictly independent of external control operations.
	
	\section{Generating Random Unitary Operations}
	\label{sec:Implementation}
	
	To enable a fair comparison between SWAP gates and random unitary operations under identical noise environments, we establish a framework in which both operations incur equivalent quantum-mechanical effort.
	This equivalence requires matching the total duration and the average energetic cost of the operations.
	The total evolution time is partitioned into $K$ discrete segments of fixed duration,
	\begin{equation}
		T_{\text{total}} = K\tau, \qquad
		\tau = \frac{\pi}{4J},
	\end{equation}
	where $\tau$ corresponds to the time needed to implement a full SWAP gate with coupling strength $J$, and $K=2L-3$ equals the number of SWAP operations in the protocol.
	During each segment $k \in {1,2,\ldots,K}$, the quantum evolution implements a unitary transformation
	\begin{equation}
		U_k(\tau) = \mathcal{T}\exp\left(-i \int_{(k-1)\tau}^{k\tau} H_k(t) dt\right) \approx R_k.
	\end{equation}
	
	The feasible unitaries $R_k$ are restricted by the quantum speed limit for unitary evolution,
	\begin{equation}
		\tau \geq \frac{\|\log U\|_F}{2 \langle \|H\|_F \rangle}
		\label{eq:general_QSL}
	\end{equation}
	where $\langle \|H\|_F \rangle$ denotes the time-averaged Frobenius norm of the Hamiltonian.
	For a SWAP gate implemented with $H_{\text{swap}} = 2J(\sigma_i^+\sigma_j^- + \sigma_i^-\sigma_j^+)$ in time $\tau = \frac{\pi}{4J}$, one finds $\|\log(\text{SWAP})\|_F = \pi J$ and $\langle \|H_{\text{swap}}\|_F \rangle = 2J$. Substituting these into Eq.~\eqref{eq:general_QSL} gives
	\begin{align}
		\frac{\pi}{4J} \geq \frac{\|\log R_k\|_F}{2 \times 2J}\implies  \|\log R_k\|_F \leq \pi J.
	\end{align}
	which is saturated for the SWAP gate.
	This bound guarantees that each $R_k$ can be physically realized within duration $\tau$; any violation would imply either a longer evolution time or a higher energetic requirement than the SWAP operation.
	
	To preserve operational equivalence, we impose a Hamiltonian-intensity constraint ensuring the same average energy expenditure as the SWAP gate:
	\begin{equation}
		\frac{1}{\tau} \int_{(k-1)\tau}^{k\tau} \| H_k(t) \|_F dt = 2J,
	\end{equation}
	where $\|H_k(t)\|_F = \sqrt{\operatorname{Tr}(H_k^\dagger(t) H_k(t))}$ quantifies the instantaneous Hamiltonian magnitude.
	This guarantees that both SWAP and random operations experience comparable noise exposure.
	
	The total unitary operation is constructed sequentially as $R^{\text{total}} = \prod_{k=1}^{K} R_k$. For each segment $k$, we first generate a candidate unitary $\widetilde{R}_k$ drawn from the Haar measure on $\text{SU}(2^L)$. To ensure physical realizability, we define the scaling factor
	\begin{equation}
		\alpha_k = \min\left( \frac{\pi J}{\|\log \widetilde{R}_k\|_F}, 1 \right),
	\end{equation}
	and rescale 
	$R_k = \exp\left( \alpha_k \log \widetilde{R}_k \right)$. The associated control fields are obtained by solving the optimization problem
	\begin{equation}
		\min \left\| R_k - \mathcal{T}e^{-i \int_0^\tau H_k(t) dt} \right\|_F \text{s.t.}  \frac{1}{\tau} \int_0^\tau \| H_k^{\text{ctrl}}(t) \|_F dt = 2J.
	\end{equation}
	After confirming that $\int_0^\tau \|H_k^{\text{ctrl}}\|_F dt = 2J\tau$ for each segment, the full operation is obtained as $R^{\text{total}} = \prod_{k=1}^{K} R_k$. 
	
	For implementation on physical transmon processors, the control amplitudes are normalized by $2J$:
	\begin{align}
		\Omega_i^k(t) &= 2J \cdot f_i^k(t), \\
		\Delta_i^k(t) &= 2J \cdot g_i^k(t), \\
		J_{ij}^k(t) &= 2J \cdot j_{ij}^k(t), \\
		K_{ij}^k(t) &= 2J \cdot k_{ij}^k(t),
	\end{align}
	subject to the normalization constraint
	\begin{equation}
		\frac{1}{\tau} \int_0^\tau \sqrt{\sum_i (|f_i^k|^2 + |g_i^k|^2) + \sum_{\langle i,j\rangle} |j_{ij}^k|^2 + \sum_{\{i,j\}} |k_{ij}^k|^2} dt = 1.
		\label{eq:norm_constraint}
	\end{equation}
	For theoretical analysis, only the optimized unitary operators $R_k$ are required, while the control-field sequences defined by Eq.~\eqref{eq:norm_constraint} provide a direct pathway to experimental realization.
	\section{Analytical origin of the fidelity optimal operation point: a two qubit example}
	\label{app:twoqubit}
	
	To elucidate the physical origin of the fidelity optimal operation point, we analyze a minimal two-qubit model undergoing two consecutive SWAP operations.
	Although such a system does not require both cavity and CPW couplings, we employ a simplified Hamiltonian to extract analytical insight into coherence behavior.
	The model Hamiltonian is defined as
	\begin{equation}
		H = \sigma_x \otimes \sigma_x + a, \sigma_y \otimes \sigma_y ,
	\end{equation}
	where superconducting-qubit frequency terms are omitted to isolate the essential structure of the noise response.
	We choose as an initial state the GHZ state, which in the two-qubit case corresponds to the Bell state,
	\begin{equation}
		|\text{GHZ}\rangle = \frac{1}{\sqrt{2}} \left( |00\rangle + |11\rangle \right).
	\end{equation}
	
	The tensor-product matrices are explicitly
	\begin{equation}
		\sigma_x \otimes \sigma_x = \begin{bmatrix} 0 & 0 & 0 & 1 \\ 0 & 0 & 1 & 0 \\ 0 & 1 & 0 & 0 \\ 1 & 0 & 0 & 0 \end{bmatrix},\
		\sigma_y \otimes \sigma_y = \begin{bmatrix} 0 & 0 & 0 & -1 \\ 0 & 0 & 1 & 0 \\ 0 & 1 & 0 & 0 \\ -1 & 0 & 0 & 0 \end{bmatrix}.
	\end{equation}
	Hence, the total Hamiltonian takes the form
	\begin{equation}
		H = \begin{bmatrix} 0 & 0 & 0 & 1-a \\ 0 & 0 & 1+a & 0 \\ 0 & 1+a & 0 & 0 \\ 1-a & 0 & 0 & 0 \end{bmatrix}.
	\end{equation}
	
\begin{figure}[t]
		\centering
		\includegraphics[width=1\linewidth]{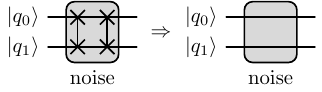}
		\caption{Quantum circuit illustrating the effect of noise on two consecutive SWAP operations in a two-qubit system.
			The noise Hamiltonian commutes with the SWAP operation, enabling a direct analysis of its influence on the overall circuit fidelity.}
		\label{fig:8}
	\end{figure}

	The SWAP operation is represented by
	\begin{equation}
		R = \begin{bmatrix}
			1 & 0 & 0 & 0 \\
			0 & 0 & 1 & 0 \\
			0 & 1 & 0 & 0 \\
			0 & 0 & 0 & 1
		\end{bmatrix},
	\end{equation}
	Since $[H, R] = 0$, the Hamiltonian commutes with the SWAP operator.
	Applying two SWAP operations yields $URUR = UURR = UU$, showing that in this configuration, the overall fidelity depends only on the noise Hamiltonian, as shown in Fig.~\ref{fig:8}. 
	The eigenvalues and eigenstates of $H$ can be directly obtained by diagonalization:
	\begin{align}
		&\lambda_1 = 1 - a, \quad \lambda_2 = -a - 1, \quad
		\lambda_3 = a - 1, \quad \lambda_4 = a + 1, \nonumber \\
		&|\phi_1\rangle = \tfrac{1}{\sqrt{2}}(1,0,0,1)^T, \quad
		|\phi_2\rangle = \tfrac{1}{\sqrt{2}}(1,0,0,-1)^T, \nonumber \\
		&|\phi_3\rangle = \tfrac{1}{\sqrt{2}}(0,1,1,0)^T, \quad
		|\phi_4\rangle = \tfrac{1}{\sqrt{2}}(0,1,-1,0)^T.
	\end{align}
	Since $|\text{GHZ}\rangle = |\phi_1\rangle$, its time evolution under $H$ is
	\begin{equation}
		|\psi(t)\rangle = e^{-iHt} |\psi(0)\rangle = \sum_i c_i e^{-i\lambda_i t} |\phi_i\rangle.
	\end{equation}
	and the fidelity is the squared overlap with the initial state,
	\begin{equation}
		F(t) = |\langle \text{GHZ} |\psi(t)\rangle|^2 = \left| \sum_i e^{-i\lambda_i t} \langle \text{GHZ} |\phi_i\rangle \cdot c_i \right|^2.
	\end{equation}

	To model noise, we perturb the Hamiltonian as
	\begin{equation}
		H = (\sigma_x \otimes \sigma_x + a, \sigma_y \otimes \sigma_y) + \delta H(t),
	\end{equation}
	where $\delta H(t)$ represents an instantaneous Gaussian perturbation with zero mean and standard deviation $\sigma$.
	The fidelity then corresponds to the ensemble-averaged overlap between the noisy and noiseless evolutions.
	For the GHZ (Bell) state, without Gaussian noise the fidelity remains unity, $F_{\text{GHZ}}(t)=1$.
	Including Gaussian fluctuations yields a Gaussian decay,
	\begin{equation}
		F_{\text{GHZ}}(t) = e^{-\sigma^2 t^2}.
	\end{equation}
	For the product-like configuration $|\uparrow\downarrow\rangle = \tfrac{1}{\sqrt{2}}(|\phi_2\rangle + |\phi_4\rangle)$, the noise-free fidelity oscillates periodically:
	\begin{equation}
		F_{\uparrow\downarrow}(t) = \cos^2(at),
	\end{equation}
	reflecting intrinsic coherence oscillations.
	Under Gaussian noise, the fidelity becomes
	\begin{equation}
		F_{\uparrow\downarrow}(t) = \cos^2(at) e^{-\sigma^2 t^2},
	\end{equation}
	demonstrating that random fluctuations induce exponential dephasing superimposed on coherent oscillations.
	
	The fidelity of these quantum states under Gaussian noise can be expressed as a function of the normalized interaction parameter $J/\lambda_0$.
	This formulation reveals how variations in SWAP coupling strength influence coherence preservation and noise susceptibility.
	While the present two-qubit model serves primarily as a qualitative illustration, it provides physical intuition that complements the numerical simulations for larger systems, where analytical treatment becomes intractable.
	\begin{figure}[t]
		\centering
		\includegraphics[width=1\linewidth]{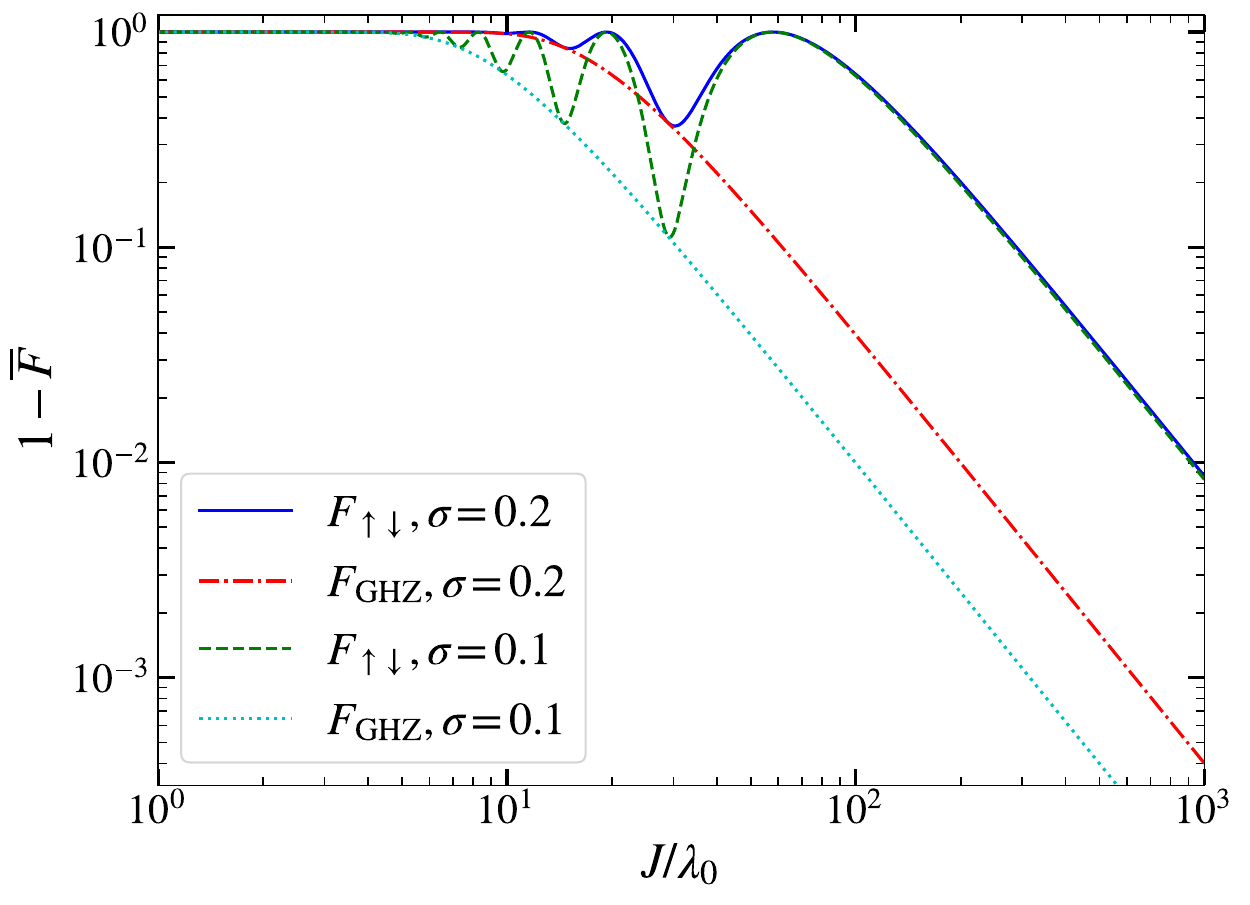}
		\caption{Fidelity of two-qubit states under Gaussian noise during two SWAP operations for different $\sigma$.
			The $|\uparrow\downarrow\rangle$ state shows a clear optimal operation point at small $J/\lambda_0$, consistent with numerical simulations.
			In contrast, the $|\text{GHZ}\rangle$ (Bell) state exhibits no optimal operation point because the simplified model omits qubit frequency terms, rendering it a Hamiltonian eigenstate and thus suppressing oscillatory evolution.}
		\label{fig:dip_analy}
	\end{figure}
	The transformation from evolution time $t$ to normalized interaction strength $J/\lambda_0$ is given by $t = \alpha \lambda_0 / J$.
	Substituting this relation into the fidelity functions yields
	\begin{equation}
		\begin{aligned}
			&F_{\text{GHZ}}\left(\frac{J}{\lambda_0}\right) = \exp\left(-\sigma^2 \left(\alpha \frac{\lambda_0}{J}\right)^2\right) \\
			&F_{\uparrow\downarrow}\left(\frac{J}{\lambda_0}\right) = \cos^2\left( \alpha \frac{a\lambda_0}{J}\right) \\
			&\times\exp\left(-\sigma^2 \left(\alpha \frac{\lambda_0}{J}\right)^2\right).
		\end{aligned}
	\end{equation}
	These expressions explicitly show how the interplay between coupling strength and Gaussian noise determines the fidelity of both quantum states.
	By setting $\lambda_{ij,0}^{t} = 10,\lambda_{ij,0}^{c} = \lambda_0$, we take $a = 10/11$.
	As shown in Fig.~\ref{fig:dip_analy}, the state $\left|\uparrow\downarrow\right\rangle$ exhibits a distinct optimal operation point at small $J/\lambda_0$, consistent with the numerical results presented in the main text.
	In contrast, the two-qubit $|\text{GHZ}\rangle$ (Bell) state shows no visible optimal operation point because, in this simplified toy model, the omission of superconducting-qubit frequency terms makes $|\text{GHZ}\rangle$ an eigenstate of the Hamiltonian, thereby suppressing oscillatory evolution.
	When these frequency terms are restored, the $|\text{GHZ}\rangle$ state is expected to exhibit an analogous optimal operation point behavior.
	
	\section{Sweet-spot dependence on interaction strength}\label{app:Jt}

	\begin{figure}[t]
		\centering
		\includegraphics[width=1\linewidth]{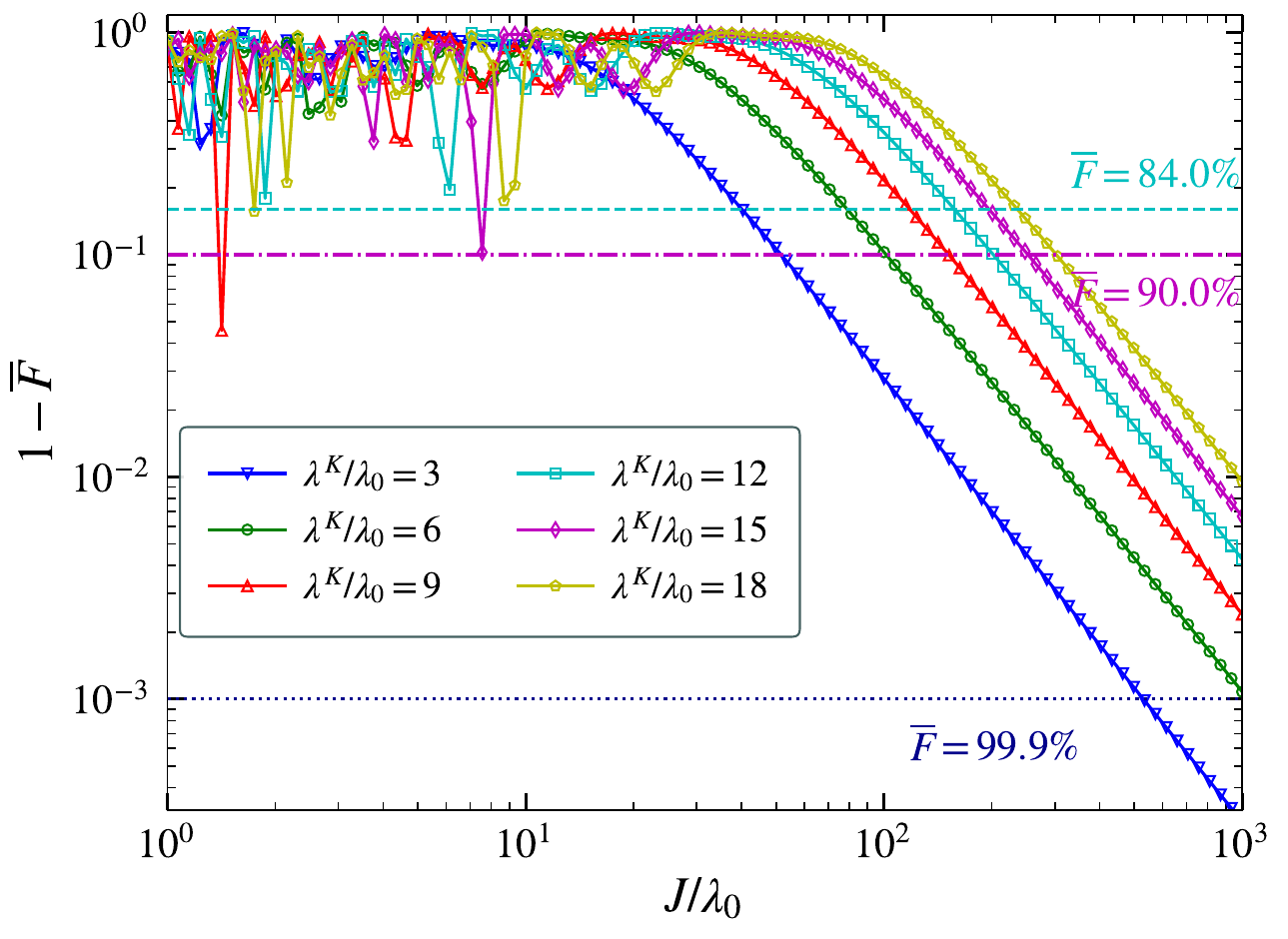}
		\caption{Average infidelity $1-\overline{F}$ of SWAP operations on $|\Psi\rangle_1$ as a function of normalized interaction strength $J/\lambda_0$ for different photon-loss coupling ratios $\lambda^{K}/\lambda_0$.
			Larger $\lambda^{K}$ corresponds to stronger noise amplitude and shifts the sweet-spot position toward smaller $J/\lambda_0$.
			Horizontal dashed lines indicate reference fidelities of $\overline{F}=84\%$, 90\%, and 99.9\%.}
		\label{fig:Jt}
	\end{figure}
	
	As discussed in Sec.~\ref{sec:Model and method}, the CPW-induced noise amplitude $\lambda_{ij}^{K}(t)$ can be three to twenty times larger than the cavity-induced term $\lambda_{ij}^{J}(t)$.
	By varying $\lambda^{K}/\lambda_{0}$ while keeping $\lambda^{J}$ fixed, we effectively tune the ratio $\lambda^{K}/\lambda^{J}$.
	Different values of $\lambda^{K}$ yield distinct sweet-spot magnitudes and corresponding pulse durations, as shown in Fig.~\ref{fig:Jt}.
	
	This variation physically corresponds to modifying the CPW parameters, with each configuration leading to a unique optimal operation time associated with its optimal operation point.
	Once the optimal parameter is identified, other unitary operations in the same device can be implemented with comparably high fidelity, consistent with the findings of Sec.~\ref{sec:generalU}.
	Interestingly, increasing $\lambda^{K}$ lowers the sweet-spot position (yielding higher fidelity) but also shortens the gate time.
	This result highlights a practical trade-off between operation duration and achievable fidelity, providing experimentally relevant guidance for system optimization.

	\section{Time-dependent noise evolution}
	\label{ssec:td_evolution}
	
	For general time-dependent noise $H_{\mathrm{noise}}(t)$, we model the total evolution using a symmetric Trotter decomposition:
	\begin{equation}
		\begin{aligned}
			U_{\mathrm{total}} \approx & \prod_{k=1}^K \Biggl( \prod_{n=1}^N 
			\exp\left(-iH_{\mathrm{noise}}(t_{k,n}) \frac{\tau}{2N} \right) 
			U_k^{1/N} \\
			& \quad \times \exp\left(-iH_{\mathrm{noise}}(t_{k,n}) \frac{\tau}{2N} \right) \Biggr)
		\end{aligned}
	\end{equation}
	where $t_{k,n} = (k-1)\tau + (n - \tfrac{1}{2}) \tfrac{\tau}{N} $ and $U_k^{1/N} = \exp\left( \tfrac{1}{N} \log U_k \right)$. This symmetric structure ensures second-order accuracy and preserves unitarity throughout the noisy evolution.

	The quantum state evolves according to the following steps:
	\begin{enumerate}
		\item Initialize $\rho_0 = |\psi_0\rangle\langle\psi_0|$.
		\item For each operation segment $k = 1, \dots, K$:
		\begin{enumerate}
			\item Set $\Delta t = \tau/N$.
			\item For $n = 1$ to $N$:
			\begin{enumerate}
				\item Apply half-step noise evolution:\\ $\rho \leftarrow e^{-iH_{\mathrm{noise}}(t_{k,n}) \Delta t/2} \rho e^{iH_{\mathrm{noise}}(t_{k,n}) \Delta t/2}$
				\item Apply the $n$th fractional unitary:\\ $\rho \leftarrow U_k^{(n)} \rho (U_k^{(n)})^\dagger$ where $U_k^{(n)} = U_k^{1/N}$
				\item Apply the second half-step noise evolution:\\ $\rho \leftarrow e^{-iH_{\mathrm{noise}}(t_{k,n}) \Delta t/2} \rho e^{iH_{\mathrm{noise}}(t_{k,n}) \Delta t/2}$
			\end{enumerate}
		\end{enumerate}
		\item Store the final density matrix $\rho_{\mathrm{final}}$.
	\end{enumerate}
	
	The fidelity reported throughout the manuscript is computed as the overlap between the final state under noisy evolution and the ideal (noise-free) final state. For a given initial state $|\psi_0\rangle$, we first compute the ideal final state $|\psi_{\text{ideal}}\rangle = U_{\text{ideal}} |\psi_0\rangle$, where $U_{\text{ideal}}$ is the target unitary (e.g., the product of SWAP gates or a general random unitary). The noisy evolution yields the final density matrix $\rho_{\text{final}}$ via the above Trotter procedure. The fidelity is then given by
	\begin{equation}
		F = \langle \psi_{\text{ideal}} | \rho_{\text{final}} | \psi_{\text{ideal}} \rangle.
	\end{equation}
	If the noisy evolution is pure (i.e., $\rho_{\text{final}} = |\psi_{\text{actual}}\rangle\langle\psi_{\text{actual}}|$), this reduces to $F = |\langle \psi_{\text{ideal}} | \psi_{\text{actual}} \rangle|^2$. This definition coincides with the standard quantum state transfer (QST) fidelity when the target operation is a state transfer protocol. In our SWAP circuits, the ideal operation cyclically permutes the qubit states so that the ideal output equals the input; thus the fidelity measures the success probability of returning the initial state, which is exactly the QST fidelity. For general random circuits, the fidelity quantifies how accurately the noisy implementation reproduces the target unitary operation, generalizing the QST fidelity to arbitrary multi-qubit transformations.

	\newpage
	
	%
	
\end{document}